\documentclass[aps,superscriptaddress,twocolumn]{revtex4-1}

\usepackage[pdftex]{graphicx}
\usepackage{dcolumn}
\usepackage{bm}
\usepackage{color}
\usepackage{amsmath}
\usepackage{nicefrac}
\usepackage{amssymb}
\usepackage[dvipsnames]{xcolor}
\usepackage{dsfont} 
\usepackage{braket}

\newcommand{\ff}[1]{{\boldsymbol #1}}

\newcommand{\bi}{\begin{itemize}}
\newcommand{\ei}{\end{itemize}}
\newcommand{\be}{\begin{equation}}
\newcommand{\ee}{\end{equation}}
\newcommand{\ba}{\begin{eqnarray}}
\newcommand{\ea}{\end{eqnarray}}

\usepackage[pdftex,colorlinks=true,linkcolor=blue,citecolor=blue,filecolor=blue]{hyperref}

\begin{document} 
  
\title{Controlling the real-time dynamics of a spin coupled to the helical edge states of the Kane-Mele model}

\author{Robin Quade}
\affiliation{I. Institute of Theoretical Physics, Department of Physics, University of Hamburg, Notkestra\ss{}e 9-11, 22607 Hamburg, Germany}

\author{Michael Potthoff}
\affiliation{I. Institute of Theoretical Physics, Department of Physics, University of Hamburg, Notkestra\ss{}e 9-11, 22607 Hamburg, Germany}
\affiliation{The Hamburg Centre for Ultrafast Imaging, Luruper Chaussee 149, 22761 Hamburg, Germany}

\begin{abstract}
The time-dependent state of a classical spin locally exchange coupled to an edge site of a Kane-Mele model in the topologically non-trivial phase is studied numerically by solving the full set of coupled microscopic equations of motion for the spin and the electron system.
Dynamics in the long-time limit is accessible thanks to dissipative boundary conditions, applied to all but the zigzag edge of interest.
We study means to control the state of the spin via transport of a spin-polarization cloud through the helical edge states.
The cloud is formed at a distant edge site using a local magnetic field to inject an electron spin density and released by suddenly switching off the injection field. 
This basic process, consisting of spin injection, propagation of the spin-polarization cloud, and scattering of the cloud from the classical spin, can be used to steer the spin state in a controlled way. 
We find that the effect of a single basic process can be reverted to a high degree with a subsequent process. 
Furthermore, we show that by concatenating several basic injection-propagation-scattering processes, the spin state can be switched completely and that a full reversal can be achieved.
\end{abstract} 

\maketitle 

\section{Introduction}
\label{sec:intro}

Topological quantum matter 
\cite{HasanKane2010,Qi2011,Ando2013,TeoKane2010,Chiu2016,Schnyder2008}
has proven to be a fascinating concept not only because of its mathematical beauty but also because it paves the way for novel applications. 
In particular, as a consequence of the bulk-boundary correspondence, quantum matter in a topologically non-trivial state is characterized by the existence of topologically protected states, which are localized at the boundaries of a sample.
These edge states are accessible by local or surface-sensitive experimental probes and feature unique physical properties. 
Their robustness against certain perturbations is highly attractive for the development of devices with new functionalities. 

Symmetries are essential for the classification of topological quantum matter \cite{Chiu2016,TeoKane2010,Schnyder2008}. 
Accordingly, the topological edge modes are distinguished by their protection against local and symmetry-preserving perturbations.
This robustness property has been exploited for various suggested applications. 

A prime example is quantum computation \cite{Kane1998,Lahtinen_2017}, employing, e.g., fractional quantum-Hall states \cite{Nayak_nonabelian_2008} or Majorana zero modes \cite{Sarma_zeromodes_2015}, where topological protection and corresponding robustness is a decisive issue.
In the field of spintronics \cite{Wolf2001}, examples are given
by one-dimensional spin transport in inverted-gap semiconductor-based devices \cite{Akhmerov2009}, 
or by spin-dependent reflection with control of the spin rotation in trilayer junctions consisting of quantum-spin Hall (QSH) and metallic materials \cite{Nagaosa2009}. 
The QSH effect can be utilized to create nearly fully spin-polarized charge currents, controlled via magnetic defects \cite{VanDyke_2016}, and all-electrical routes have been suggested to manipulate the spin of a magnetic adatom at the edge of a QSH insulator \cite{Narayan_2013,Hurley_Narayan_Aaron_2013}.

Our present study addresses the helical edge states of a time-reversal (TR) symmetric two-dimensional topological insulator and concentrates on the Kane-Mele (KM) model as a prototype \cite{KaneMele2005,KaneMele2005Z2}. 
Topologically non-trivial properties of the KM model originate from its TR symmetric spin-orbit coupling term.
In particular, this induces the QSH effect.
The model has originally been proposed for graphene \cite{KaneMele2005} but turned out to be more relevant for quantum-well systems \cite{Bernevig1757,Koenig766}.
It can also be understood to describe a class of graphene-like two-dimensional monolayer honeycomb materials that feature significant spin-orbit interaction, such as silicene and related systems \cite{Liu2011_LEH,Ezawa2015}.
It has recently been shown that an interacting KM model emerges as an effective low-energy theory in stacked 1T-TaSe2 bilayers \cite{PAZ+20}.
Quite generally, the KM model represents a paradigmatic model for two-dimensional class-AII topological insulators with phases characterized by a $\mathbb{Z}_{2}$ index and protected by time-reversal symmetry (TRS).

It is quite natural to probe TRS protected topological states of $\mathbb{Z}_{2}$ insulators by means of TRS breaking local perturbations \cite{chang_wei_moodera_2014}.
This idea has been pursued in various studies of TR symmetric topological systems, such as Bi$_2$Te$_3$, Bi$_2$Se$_3$, Sb$_2$Te$_3$, by doping with magnetic transition-metal atoms \cite{Yu2012,Chen2010} or by depositing magnetic adatoms at the surface \cite{EWS+14,VCM+16}.
Locally breaking TRS may lead to rather exotic phenomena such as an image magnetic monopole \cite{Qi2009}.

Especially interesting is the interaction between two magnetic adatoms that is mediated by the helical edge states. 
In the limit of a weak exchange coupling $J$ between adatoms and substrate, standard concepts of RKKY theory \cite{Kit68} can be adapted to tight-binding or continuum models for helical QSH boundary states.
In the vicinity of a classical magnetic impurity, the local (spin) density of states is suppressed at low energies \cite{LLX+09}. 
The RKKY interaction between two impurities becomes ferromagnetic for a chemical potential such that the Fermi wavelength is much larger than the impurity-impurity distance. 
In general the coupling is noncollinear and in-plane with a power-law spatial decay \cite{GCXZ09}.
There is an additional \cite{KKB17} Bloembergen-Rowland-type \cite{BR55} bulk contribution decaying exponentially with distance.
A weakly broken TRS gaps out the Dirac cone and induces a strongly anisotropic RKKY coupling with Dzyaloshinsky–Moriya and with  both, in-plane and out-of-plane Ising terms, decaying exponentially with the distance between the impurities for a chemical potential within the gap \cite{HKD20}.
Effects of strong electron interaction and the Doniach competition of indirect exchange with Kondo screening in a helical Luttinger liquids limit the applicability of RKKY theory \cite{YY18}. 

The {\em real-time dynamics} of magnetic impurities at surfaces of topological insulators has been studied to a lesser extent.
Recent studies have employed time-dependent density-functional theory \cite{BDGL19} and, for periodically driven impurities, Floquet theory \cite{PF19}. 
The effects of a single TRS-breaking magnetic impurity on the transport properties of a KM zigzag ribbon mediated by the helical edge states have been studied within scattering theory \cite{Tatsumi_etal_2020}.
The long-time dynamics of a single classical spin exchange coupled to the edge of a Su-Schrieffer-Heger model has been investigated recently \cite{EP21}.
Apart from the latter, the full microscopic real-time dynamics beyond the linear-response approach \cite{BN19,ON06,BNF12} has not yet been addressed so far.

In the present work we numerically study the real-time dynamics of a classical ``read-out'' spin that is exchange coupled to the local spin of the electron system at a site of a zigzag edge of the KM model on a honeycomb lattice. 
The dynamical state of the read-out spin is affected by another impurity at a distant site on the same edge, which is used to inject a local spin excitation. 
The transport of the injected spin density through the helical edge state and its impact on the classical spin is microscopically traced as function of time. 
Note that local breaking of TRS due to the read-out impurity spin enables backscattering \cite{Maciejko2009,Tanaka2011} of the transported spin density and thus allows driving of the read-out spin dynamics.
Our goal is to fully control the state of the classical spin by local excitations of the system over long distances making use of the topological protection of the edge state.
We demonstrate that this can be achieved with a predefined precision by iterating the spin-injection and transport process.

The studied setup is partially motivated by the further progress of experimental techniques, e.g., in detecting states of magnetic adatoms \cite{Wie09} and measuring indirect magnetic (RKKY) interactions on a nanoscale \cite{ZWL+10}. 
Scanning-tunnelling microscopy with two or several tips \cite{XTD06,KOZ+17,HZN+17,VCK+18}, with a separation down to the nanometers, ideally with magnetic tips and spin resolution, would be perfectly suited to initiate, probe and control spin-momentum-locked transport. 
Our study aims at an improved understanding of manipulation of local magnetic states through topological surface states in the time domain.
As time-dependent STM techniques \cite{vHZ10,YYGJ18} address the $\mu$s rather than the ps time regime, however, we focus on the initial and the final-state spin configurations. 

A numerically exact solution of the coupled set of equations of motion for the classical read-out spin and for the dynamics of the entire electronic system can only be achieved for a lattice of finite size. 
Here, we demonstrate that a ribbon-shaped geometry with only 4 unit cells in the direction perpendicular to the zigzag edge and with about 100 sites along the zigzag edge is fully sufficient to achieve propagation times of the order of $10^{3}$ inverse nearest-neighbor hoppings (and more) and thus allows for a complete monitoring of the injection-transport-driving process.
This, however, requires special boundary conditions \cite{EP20} for the armchair edges and for the opposite zigzag edge.
Namely, the boundaries must fully absorb the residual propagating excitations to avoid reflections. 
As has been shown recently for a spin coupled to a one-dimensional topological chain \cite{EP21}, this can be achieved with properly modified Lindblad-type edge potentials.
This scheme is adapted here and applied to the considered ribbon geometry. 
It allows us to easily achieve the required propagation times without any unwanted interference effects and without any modification of the physical real-time dynamics in the core of the system.

The rest of the paper is organized as follows: 
The next sections \ref{sec:mod} and \ref{sec:dyn} introduce the model and our approach to compute time-dependent observables.
In section \ref{sec:rib} we specify the geometry in detail and discuss a static spin injection and the subsequent propagation of the spin-polarization cloud.
Scattering of the polarization cloud and its impact on the read-out spin are addressed in Secs.\ \ref{sec:read} and \ref{sec:impact}.
A basic process that consists of a dynamic spin-injection and subsequent pumping of the read-out spin can be reverted or iterated, as described in Sec.\ \ref{sec:iter}. 
Section \ref{sec:con} summarizes our findings and gives a brief outlook.

\section{Kane-Mele $s$-$d$ model}
\label{sec:mod}

Using standard notations, the Hamiltonian of the Kane-Mele model \cite{KaneMele2005,KaneMele2005Z2} is given by:
\begin{gather}
	H_{\rm KM} = t_{\rm hop} \sum_{\braket{i,j},\alpha} c_{i\alpha}^{\dagger}  c_{j\alpha} 
	+ i t_{\text{so}} \sum_{\braket{\braket{i,j}},\alpha,\beta} \nu_{ij} 
	c_{i\alpha}^{\dagger}  \sigma_{\alpha \beta}^{(z)}  c_{j\beta}  
	\nonumber \\ 
	+ V \sum_{i,\alpha} \epsilon_{i}  c_{i\alpha}^{\dagger}  c_{i\alpha}
	\: .
	\label{eq:hkm}
\end{gather}
Here, $i$ and $j$ label the sites of a honeycomb lattice, $\alpha, \beta = \uparrow, \downarrow$ is the spin projection, $\braket{. \,,.}$ and $\braket{\braket{.\,,.}}$ indicate summation over nearest or next-nearest neighbors, respectively, and $\sigma^{(z)}$ is the $z$-Pauli matrix.
The spin-diagonal nearest-neighbor hopping amplitude $t_{\rm hop} \equiv 1$ sets the energy unit and with $\hbar \equiv 1$ also the time unit. 
The anisotropic spin-orbit coupling (SOC), responsible for the helicity of the topological boundary states, is modeled as a spin-dependent next-nearest neighbor hopping with a sign factor $\nu_{ij} = \pm 1$, which is positive (negative) for anticlockwise (clockwise) hopping $j \to i$ within a hexagon of the lattice.
The strength of the SOC is controlled by the amplitude $it_{\rm so}$ which is purely imaginary ($t_{\rm so} >0$). 
In combination with the sign factor this ensures hermiticity of the SOC term. 
Finally, $V>0$ is the strength of a sublattice-parity-breaking ionic potential, which includes a sign $\epsilon_{i}= +1(-1)$ if site $i$ belongs to sublattice $A$ ($B$).
The $V$ term is used to tune the ground state of the system between topologically distinct phases. 
For given $t_{\rm so}$, we choose 
\be
V = 3 \sqrt{3} t_{\rm so} \pm 0.5 \Delta
\label{eq:gap}
\ee 
to produce a band structure with a desired (small) bulk band gap $\Delta>0$.
For positive (negative) sign this yields the topologically trivial (non-trivial) phase. 
This choice of $V$ has the convenient feature that the bulk gap is the same in both phases.

Each term of the KM model is invariant under TR, i.e., its Hamiltonian commutes with the antiunitary TR operator ${\Theta} = e^{- i \pi {S}_{y}}  {\mathcal{K}}$ where ${S}_{y}$ is the $y$-component of the total spin and ${\mathcal{K}}$ is the complex conjugation with respect to the spinful orbital basis, such that ${\Theta} c_{i\alpha}^{(\dagger)} {\Theta} = i \sum_{\beta} \sigma_{\alpha\beta}^{(y)} c_{i \beta}^{(\dagger)}$.
In the single-electron subspace, ${\Theta}$ squares to $-{\mathds{1}}$.
This entails Kramers degeneracy of the single-particle eigenstates of $H_{\rm KM}$.

The two topologically distinct bulk phases of the KM model are distinguished by a topological $\mathbb{Z}_{2}$ invariant $\nu$, with $\nu = 0$ and $\nu =1$ referring to the topologically trivial and nontrivial phases, respectively.
It can be defined as \cite{KaneMele2005,KaneMele2005Z2}
\begin{equation}
	(-1)^{\nu} = \prod_{\Gamma_{i} \in \ff \Gamma} \frac{\sqrt{\det \omega(\Gamma_{i})}}{\text{Pf }\omega(\Gamma_{i})}
\end{equation}
in terms of the set of TR symmetric invariant momenta $\ff \Gamma$ and the $2 \times 2$ TRS ``scattering'' matrix \cite{Kaufmann_2016}
\begin{equation}
	\omega_{rs}(\Gamma_{i}) = \braket{u_{r}(\Gamma_{i})|\Theta|u_{s}(\Gamma_{i})}
	\: .
\end{equation}
where the indices $r,s=1,2$ label the two Bloch states $\ket{u_{1}(\mathbf{k})}$, $\ket{u_{2}(\mathbf{k})}$ of the occupied Kramers pair.
The bulk-boundary correspondence of the KM model is expressed as $\nu = N_{\rm K} \; \text{mod } 2$, where $N_{\rm K}$ is the number of Kramers boundary pairs, i.e., the number is even (odd) in the topologically trivial (non-trivial) phase.
The topological edge states of the model are helical, meaning that the two spin species are interlocked with opposite momenta and hence opposite propagation directions.
Note that helical edge states are ``immune'' to TRS-preserving perturbations but scatter from TRS-breaking perturbations \cite{Maciejko2009,Tanaka2011}.

We consider the KM model in a nanoribbon geometry with topological zigzag edges (see Sec.\ \ref{sec:rib} below). 
At a site $R$ of a zigzag edge, a classical (``read-out'') spin $\ff S_{\rm R}$ of fixed length is exchange coupled to the local spin $\ff s_{\rm R} = \frac{1}{2} \sum_{\alpha\beta} c_{{\rm R}\alpha}^{\dagger} \ff \sigma_{\alpha\beta} c_{{\rm R}\beta}$ of the electron system. 
Here, $\ff \sigma$ is the vector of Pauli matrices.
This perturbation, $J \ff S_{\rm R} \ff s_{\rm R}$, locally breaks TRS opposed to, e.g., a Kondo coupling to a quantum-spin $1/2$. 
The classical spin is a proper way to describe a ``magnetic'' adatom with a well-formed spin moment that is stable on a time scale exceeding all other relevant time scales of the system. 

The classical spin $\ff S_{\rm R}$ is susceptible to a spin-density excitation propagating along the edge through gapless helical edge states, the presence of which is enforced in case of a topologically non-trivial phase of the bulk electron system.
Local breaking of TRS allows the spin excitation to scatter from the impurity spin and thereby to exchange a spin torque that drives the dynamics of $\ff S_{\rm R}$.
Due to the topological protection and due to the helicity of the edge states the electron-spin density is transported robustly and unidirectionally.

Locally initializing a spin excitation in the electron system that is confined to the zigzag edge likewise requires another TRS-breaking local perturbation. 
This could be achieved by means of a second magnetic adatom with a spin moment that is externally driven, e.g., by means of a spin-polarized STM tip.
For simplicity, we will here model the spin-injection process by a local magnetic field $\ff B_{\rm I}$ 
which couples to the local electron spin at a site $I$ and which is suddenly switched on and off to induce real-time dynamics.

The Hamiltonian of the total system, consisting of the Kane-Mele electron model and the two local perturbations, the exchange-coupled read-out spin $\ff S_{\rm R}$ at edge site $R$ and the local magnetic field $B_{\rm I}$ at edge site $I$, then reads:
\begin{equation}
	H = H_{\rm KM} + J \ff{S}_{\rm R}\ff{s}_{\rm R} - \ff{B}_{\rm I}\ff{s}_{\rm I} \: .
\label{eq:hfull}	
\end{equation}
$J$ is the strength of the local exchange interaction. 
$H$ represents a two-impurity $s$-$d$-type model \cite{Z51,V+} with a possibly non-trivial bulk topological ground state. 

\section{Real-Time Dynamics}
\label{sec:dyn}

Our ambition is to fully determine the microscopic real-time dynamics of the system, Eq.\ (\ref{eq:hfull}).
To this end we introduce the one-particle reduced density matrix $\ff \rho$ with elements
$\rho_{i\alpha,j\beta} = \langle \Psi(t) | c_{j\beta}^{\dagger} c_{i\alpha} | \Psi(t)) \rangle$, where $|\Psi(t)\rangle$ is the $N$-particle quantum state of the electron system at time $t$ which implies $\mbox{tr} \ff \rho = N$.
We consider a half-filled system, where $N = L$ is given by the number of lattice sites $L$. 
For the quantum-classical hybrid system \cite{Elz12,Yang_1980}, there is a closed system of equations of motion \cite{SP15}, consisting of a Landau-Lifschitz-type equation, 
\begin{equation}
\frac{d\ff S_{\rm R}}{dt} = J \braket{\ff s_{\rm R}} \times \ff S_{\rm R} \: ,
\end{equation}
with $\langle \ff s_{\rm R} \rangle = \frac{1}{2} \sum_{\alpha\beta} \rho_{R \alpha, R \beta} \ff \sigma_{\beta\alpha}$
for the read-out spin and a von Neumann-type equation,
\begin{equation}
i \frac{d \ff \rho}{dt} = [\ff T_{\rm eff} , \ff \rho] \: ,
\label{eq:vn}
\end{equation}
for the density matrix. 
Here, $\ff T_{\rm eff}$ is the effective hopping matrix with elements given by 
\be
T_{\rm eff, i\alpha, j\beta} = T_{i\alpha, j\beta} + \delta_{iR}\delta_{jR} J \frac{1}{2} \ff \sigma_{\alpha\beta} \ff S_{\rm R}
- \delta_{iI}\delta_{jI} \frac{1}{2} \ff \sigma_{\alpha\beta} \ff B_{\rm I} 
\: ,
\ee
and $\ff T$ is the bare hopping matrix of the unperturbed Kane-Mele model, Eq.\ (\ref{eq:hkm}):
\begin{gather}
T_{i\alpha,j\beta} = t_{\rm hop} \delta_{\braket{i,j}} \delta_{\alpha \beta} + i t_{\rm so} \delta_{\braket{\braket{i,j}}} \nu_{ij} \sigma_{\alpha \beta}^{(z)} + V \epsilon_{i} \delta_{ij} \delta_{\alpha \beta} \: .
\end{gather}
Using standard methods for systems of ordinary differential equations, this can be solved conveniently for finite systems with up to $L \approx 10^{3}$ sites \cite{SP15,SP17}.

\begin{figure}[t]
\includegraphics[width=0.8\columnwidth]{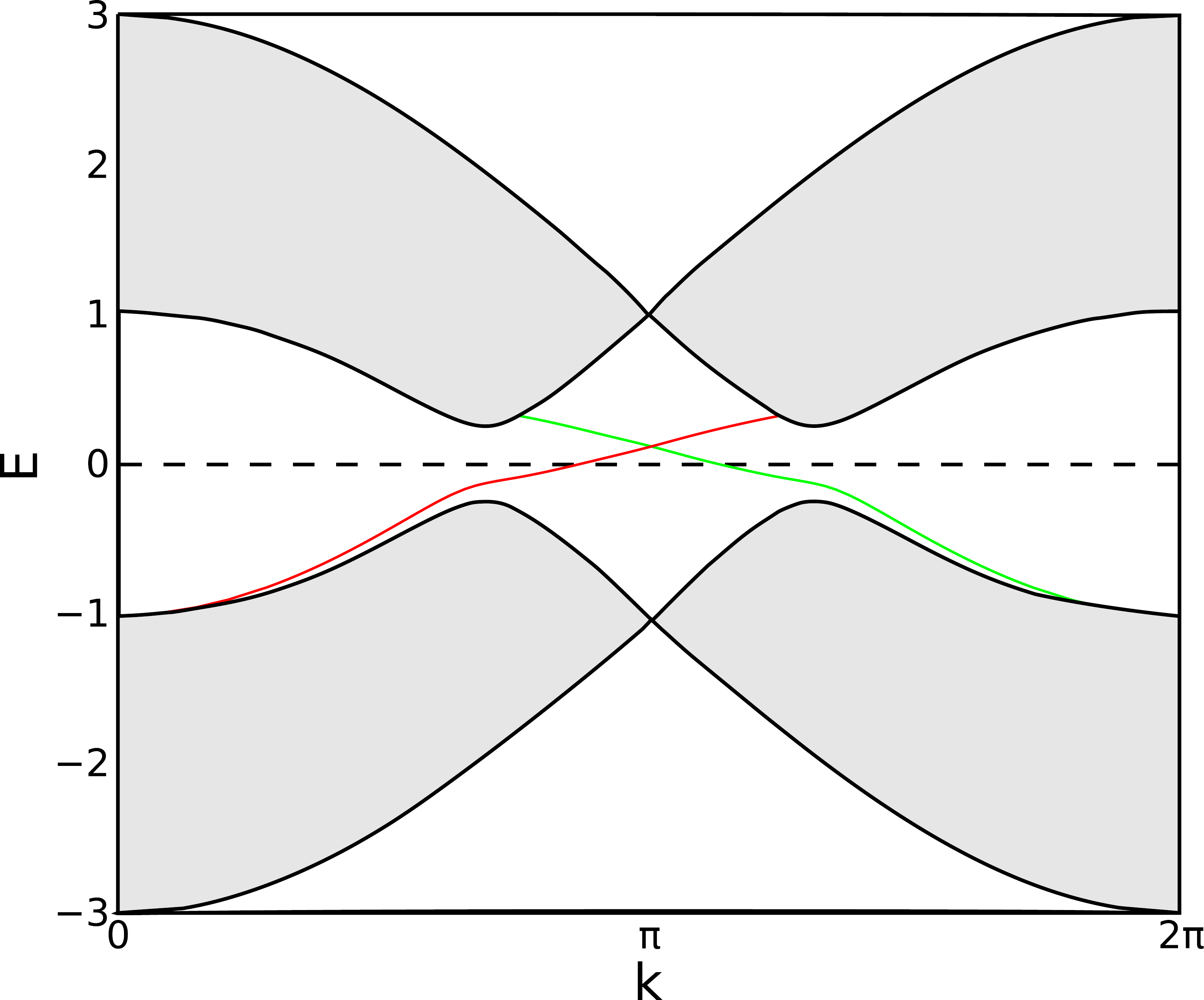}
\caption{
Bandstructure of the Kane-Mele model, Eq.\ (\ref{eq:hkm}), projected onto a zigzag edge.
Gray area: 
projected bulk bands as obtained from a calculation for a ribbon geometry with 20 unit cells in the normal direction.
Red (green) line: 
dispersions $\varepsilon_{\uparrow}(k)$ ($\varepsilon_{\downarrow}(k)$) of the two helical edge states outside the bulk continuum (edge states localized at the opposite edge are not displayed).
$t_{\rm hop} = 1$ sets the energy scale. 
Calculation for $t_{\rm so} = 0.05$, band gap $\Delta = 0.3$ and $V$ as obtained from Eq.\ (\ref{eq:gap}) for the negative sign.
}
\label{fig:bands}
\end{figure}

Initiating the real-time dynamics by a local time-dependent perturbation, causes excitations propagating as a wave packet -- predominantly along the edge. 
The propagation speed is roughly given by the Fermi velocity $v_{\rm F} = d \varepsilon_{\sigma}(k=k_{\rm F}) / dk$ of the two edge states, where $\varepsilon_{\rm F}=0$ is the Fermi energy.
Fig.\ \ref{fig:bands} gives an example. 
It displays the bulk bandstructure of the unperturbed model, Eq.\ (\ref{eq:hkm}), projected onto a zigzag edge.
Parameters (see caption) are chosen such that the topologically non-trivial phase is realized. 
We can read off $v_{\rm F} \approx \pm 0.285$ in good agreement with $v_{\rm F} \approx \pm 0.286$ as obtained from an analytical expression given in Ref.\ \cite{Cano_Cortes_2013}

Addressing the full microscopic real-time dynamics requires a system of finite (but large) size.
In case of open boundary conditions, an initially excited wave packet propagating along the edge with velocity $\pm v_{\rm F}$ at a distance $d$ from one of the corners will be reflected after a time $t_{\rm refl.} \approx d / |v_{\rm F}|$, back-propagate, and finally lead to interferences at the position of the impurity spin. 
Such unwanted finite-size effects can be suppressed almost completely with the help of absorbing boundary conditions that have been introduced and tested in Ref.\ \cite{EP20}.
This allows us to study real-time dynamics on long time scales even for moderate system sizes, see Sec.\ \ref{sec:iter}. 
Following Ref.\ \cite{EP20} we therefore replace Eq.\ (\ref{eq:vn}) by 
\begin{equation}
\frac{d \ff \rho(t)}{dt}  = -i [\ff T_{\rm eff}(t),\ff \rho(t)]  - \{ \ff \gamma , \ff \rho(t) - \ff \rho(0)\} \; ,
\label{eq:lin}
\end{equation}
where $\{\cdot, \cdot\}$ denotes the anticommutator.
This equation contains an additional boundary term involving a diagonal and non-negative matrix $\ff \gamma$ with 
elements
\begin{equation}
	\gamma_{i \alpha, j\beta} = \delta_{ij} \delta_{\alpha \beta} \gamma_{i \alpha} 
	\: .
\label{eq:lind}
\end{equation}
$\gamma_{i \alpha}$ is nonzero for sites $i$ in a thin ``Lindblad shell'' at the system boundaries, except, of course, for the ``physical'' zigzag edge of interest.
In our case, see Fig.\ \ref{fig:rib1}, the sites of the Lindblad shell are those of the three (``unphysical'') edges of the ribbon coupling to the Lindblad bath shown in red color.

In fact, Eq.\ (\ref{eq:lind}) is derived by starting from the general Lindblad equation \cite{Lindblad_1976,Pearle_2012} but allowing only sites in the Lindblad shell close to the boundaries to couple to the environment. 
In a next step the Lindblad theory is specialized to a system of noninteracting electrons. 
This results in an equation of motion involving the one-particle reduced density matrix only, rather than the many-body statistical operator. 
Finally, and most importantly, an additional term $\propto \rho(0)$ is incorporated which is missing in standard Lindblad-based computations and which is necessary to avoid artificial excitations {\em initially generated} due to the coupling to the bath. 
With a standard Lindblad bath extended by this extra term, reflections of physical excitations from the boundaries as well as initial-state artifacts can be suppressed very efficiently.
This has been demonstrated recently \cite{EP20,EP21} to work well for one-dimensional tight-binding systems with classical-spin impurities.
The approach describes time evolution respecting total-probability conservation. 
Energy and angular momentum (spin), on the other hand, are only conserved locally but dissipated at the boundaries.

Here, we adapt the method to a two-dimensional nanoribbon geometry to construct absorbing boundaries on the zigzag edge opposite to the edge of interest as well as on the armchair boundaries.
For the geometrical setup and the propagation time scales discussed in the sections below, it has turned out that a uniform shell of unit thickness and a spin-independent $\ff \gamma$-matrix is fully sufficient.
This means that only a single scalar parameter $\gamma$ must be fixed. 
This parameter controls the rate of dissipation at the absorbing boundaries and is set $\gamma= 0.2$ throughout this study, following Refs.\ \cite{EP20,EP21}.
By comparing, for sufficiently short time scales, with the open-boundary dynamics [Eq.\ (\ref{eq:vn})], we have carefully checked that the presence of absorbing Lindblad sites with $\gamma=0.2$ does not affect the physical dynamics in the core region of interest, i.e., on the zigzag edge to which the impurity terms in Eq.\ (\ref{eq:hfull}) are coupled.

\section{Ribbon geometry, spin injection and propagation}
\label{sec:rib}

\begin{figure}[t]
\includegraphics[width=0.95\columnwidth]{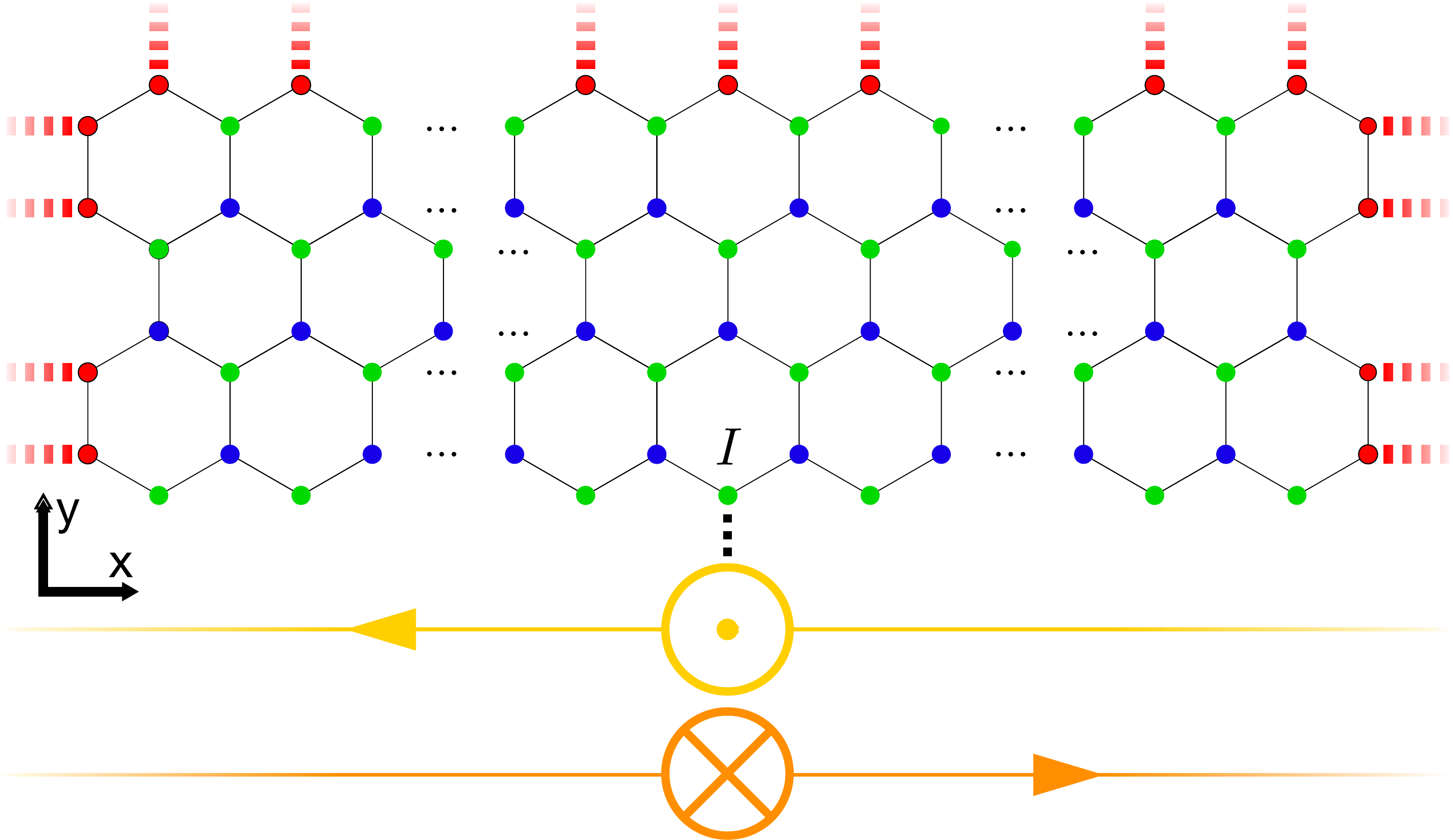}
\caption{
Sketch of the Kane-Mele model in a ribbon geometry.
Green (blue) dots: sites of sublattice A (B). 
Red dots: edge sites with coupling to a Lindblad bath (thick dashed red lines).
Yellow (orange) symbol $\odot$ ($\otimes$): 
out-of-plane magnetic field $\ff B_{\rm I}$ in positive (negative) $z$-direction coupled (dotted black line) to an ``injection site'' $I$.
Here, $I$ is chosen as the central site of the ``physical'' zigzag edge (without Lindblad boundary).
A finite $\ff B_{\rm I} = + |B_{\rm I}| \ff e_{z}$ ($\ff B_{\rm I} = - |B_{\rm I}| \ff e_{z}$) induces a finite spin-up (spin-down) $z$-polarized spin density.
In the spin-up case, the excitation propagates to the left (yellow line with arrow at the bottom), once the injection field is switched off.
Vice versa, a spin-down polarized density propagates to the right (orange line).
}
\label{fig:rib1}
\end{figure}

We consider the Kane-Mele model in a ribbon geometry as shown in Fig.\ \ref{fig:rib1}.
The numerical effort for computing the full real-time dynamics of the electronic system asymptotically scales as $L^{2}$ for a large total number of lattice sites $L$.
Computations for systems with $L \approx 500$ are convenient. 
For the  calculations we choose a ribbon geometry with $L=572$ sites. 
The short side of the ribbon along the armchair ($y$-) direction extends over four unit cells consisting of two sites each, while along the $x$-direction the ribbon extends over 71.5 unit cells, as displayed in Fig.\ \ref{fig:rib1}.

In the figure, the lower zigzag edge is the physical edge. 
Sites at the remaining three edges are coupled to the Lindblad bath, as discussed in Sec.\ \ref{sec:dyn} above and as shown in red color. 
Considering local time-dependent perturbations at the physical edge, this setup almost perfectly simulates the real-time dynamics of a half-infinite sample.
Note that the helical edge states are exponentially localized on the zigzag edge.
We have checked that with four unit cells in $y$-direction, the overlap of edge states with edge states on the opposite zigzag edge, which are perturbed by the Lindblad boundary condition, is in fact negligible. 
Furthermore, on the physical edge, the edge states are predominantly localized on the A-sublattice sites, such that it is advantageous to couple the two B-sublattice edge sites to the Lindblad bath.
This motivates the choice of an extra half unit cell in $x$-direction.
At the other edges both, B- and A-sublattice sites, are coupled to the Lindblad bath.

Evidently, a larger system generally comes with reduced finite-size artifacts. 
We have verified that the results for the real-time dynamics presented in the following are robust against both, shape-preserving and shape-altering rescaling of the geometry, mainly by comparing with results obtained for smaller $L$.

We start the discussion of the results with a static spin injection process. 
To this end, we consider the Kane-Mele ribbon with an additional local magnetic field $\ff B_{\rm I}$ pointing out-of-plane, i.e., in $+z$- or ($-z$)-direction, see yellow (orange) $\odot$ ($\otimes$) symbols in Fig.\ \ref{fig:rib1}, and coupling locally to an ``injection site'' $I$ of the physical zigzag edge.
We first choose $I$ as the central site on this edge.

\begin{figure}[t]
\includegraphics[width=0.98\columnwidth]{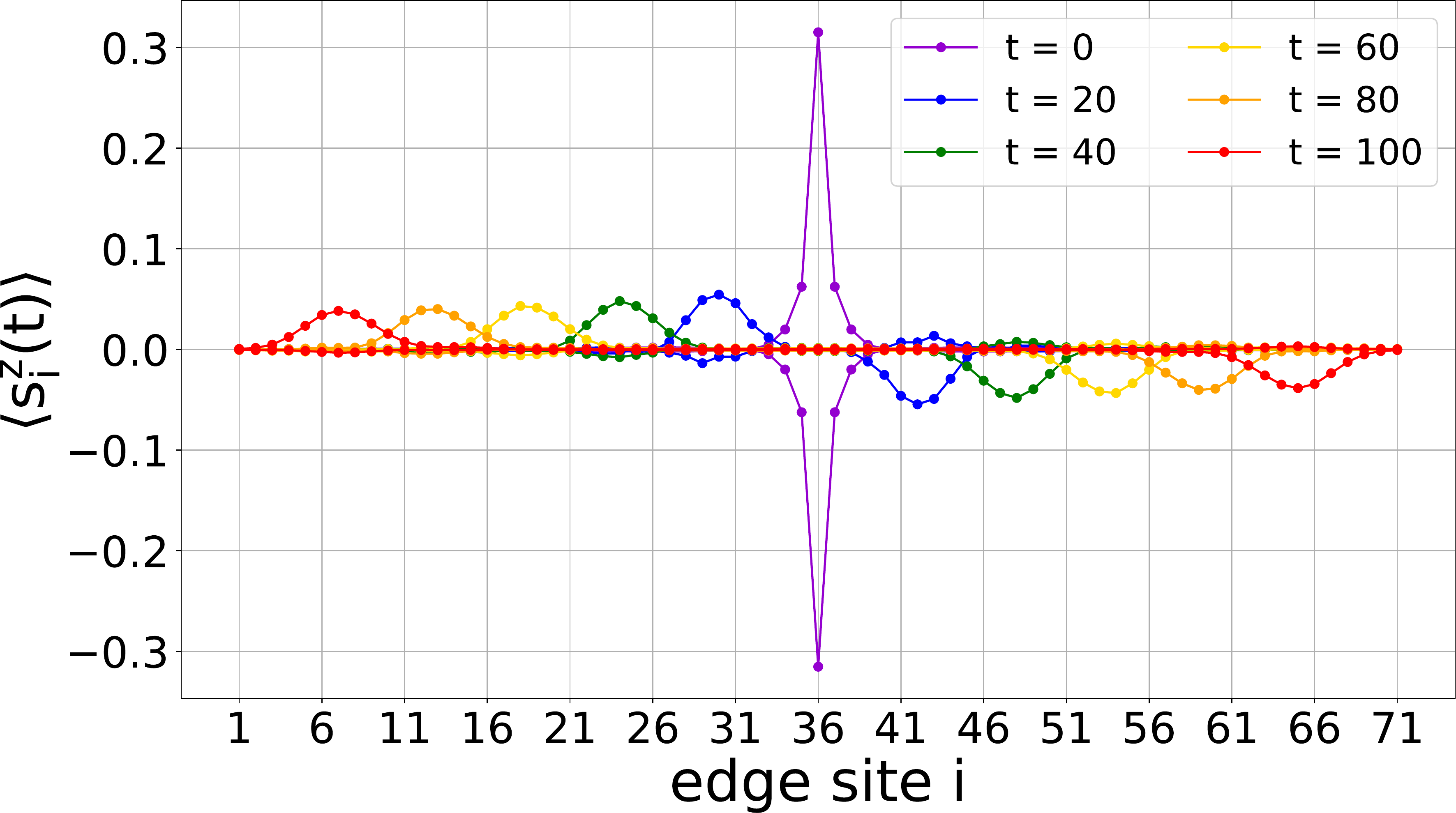}
\caption{
Snapshots of the spatial distribution of spin-polarization $\braket{{s}_{i}^{z}(t)}$ at selected instants of time (colors).
At $t=0$ the spin-injection field $\ff B_{\rm I}$ at the central site $I=36$ of the zigzag edge ($A$-sublattice edge sites are enumerated from $i=1$ to $i=71$) has suddenly been been switched off and the excitation is released.
Upper part with $\braket{{s}_{i}^{z}(t)}>0$: spin-up injection. 
Lower part with $\braket{{s}_{i}^{z}(t)}<0$: spin-down injection. 
The time unit is set by the inverse hopping parameter $t_{\rm hop}^{-1}=1$ ($\hbar\equiv 1$).
Further parameters: $t_{\rm so}=0.05$, $\Delta=0.3$ (as in Fig.\ \ref{fig:bands}), $V$ as obtained from Eq.\ (\ref{eq:gap}) with negative sign, initial field strength $B_{\rm I}=1$, Lindblad parameter $\gamma = 0.2$. 
}
\label{fig:rel}
\end{figure}

The pure Kane-Mele nanoribbon without the injection field, Eq.\ (\ref{eq:hkm}), is time-reversal symmetric. 
Without the SOC term, the Hamiltonian is additionally invariant under global $SU(2)$ rotations generated by the total electron spin. 
The SOC introduces an anisotropy and reduces this symmetry to a $U(1)$ rotational symmetry of the Kane-Mele Hamiltonian around the $z$-axis. 
The ground state, however, is an unpolarized Fermi sea of the form $| \mbox{g.s.} \rangle = \prod_{k\le k_{\rm F}} c^{\dagger}_{k\uparrow} c^{\dagger}_{k\downarrow}| \mbox{vac.}\rangle$, since the total-spin $z$-component is conserved. 
Thus, the ground state is a non-degenerate SU(2)-symmetric spin singlet. 
Any additional local magnetic field $\ff B_{\rm I}$ reduces the SU(2) symmetry of the ground state to a U(1) symmetry, but the ground-state energy is independent of the direction of the field.
Contrary, the U(1) symmetry of the Hamiltonian is broken by any field with a nonzero $z$-component.
Here, we choose $\ff B_{\rm I} = \pm B_{\rm I} \ff e_{z}$. 
In this case, we have a $U(1)$ spin-rotation invariance around the $z$-axis for both, the Hamiltonian and its ground state.
A field in $z$-direction induces a spin-up (spin-down) polarization of the local magnetic moments in the vicinity of site $I$, which we expect to show unidirectional (spin-momentum-locked) propagation.

Calculations are performed for $t_{\rm hop}=1$, $t_{\rm so}=0.05$, $\Delta=0.3$, and with $V$ fixed by Eq.\ (\ref{eq:gap}) for the topologically non-trivial case, as in Fig.\ \ref{fig:bands}. 
The field strength is set to $B_{\rm I}=1$ and the Lindblad parameter to $\gamma = 0.2$.
The resulting ground-state local moment at site $I$ amounts to $\braket{{s}_{I}^{z}(t=0)} \approx 0.33 < 0.5$, i.e., the local moment is not fully polarized.
Fig.\ \ref{fig:rel}, for time $t=0$, shows the entire ground-state polarization cloud induced by $\ff B_{\rm I} = + B_{\rm I} \ff e_{z}$ (see upper half, purple data). 
Data for $\ff B_{\rm I} = - B_{\rm I} \ff e_{z}$ (shown in the lower half) differ by the sign of  $\braket{{s}_{i}^{z}(t=0)}$ only.

\begin{figure}[t]
\includegraphics[width=0.98\columnwidth]{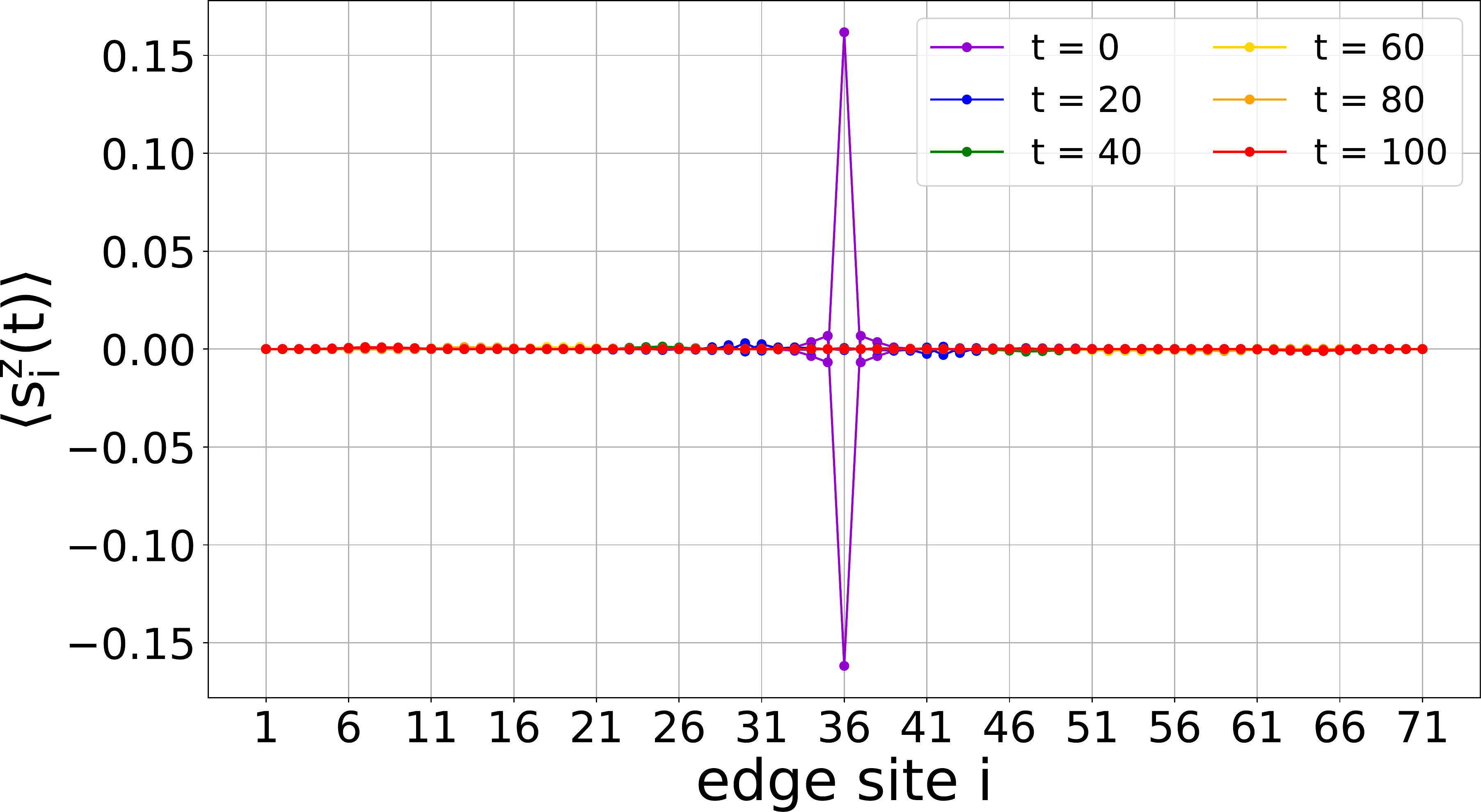}
\caption{
The same as Fig.\ \ref{fig:rel} but for the topologically trivial case with $V = 3 \sqrt{3} t_{\rm so} + 0.5 \Delta$, see Eq.\ (\ref{eq:gap}).
}
\label{fig:triv}
\end{figure}

To release the spin excitation, we suddenly switch off the field at time $t=0$, i.e., $\ff B_{\rm I}=0$ for $t>0$.
Right after the injection, at $t=0$, the polarization cloud is spread over about 5 sites in both cases. 
In the course of time, it continuously broadens and spreads over about 10 sites at $t=100$ inverse hoppings after the injection.
The results for spin-up and spin-down injection are completely identical apart from the important fact that the spin-up excitation mainly propagates to the left, while the spin-down excitations moves to the right along the zigzag chain. 
This is exactly the effect of spin-momentum locking in the topologically non-trivial state.
We also note that at early times (of the order of about 10 inverse hoppings) the total weight of the cloud is approximately halved.
This is attributed to the fact that about a half of the injected spin density is carried by bulk states and thus immediately transported away from the edge and dissipated into the bulk. 

This interpretation is corroborated by a calculation for a different strength of the ionic potential, where in Eq.\ (\ref{eq:gap}) the positive sign is used, i.e., $V = 3 \sqrt{3} t_{\rm so} + 0.5 \Delta$. 
This yields the same size of the bulk gap $\Delta$ but the topologically trivial state.
Fig.\ \ref{fig:triv} displays the according results.
In this case one finds a much less developed polarization cloud with about a factor of two smaller local moment $|\braket{{s}_{I}^{z}(t)}| = 0.16$ right after the injection. 
The propagation of both, a spin-up and a spin-down excitation, is completely identical. 
The polarization cloud at the edge sites does not propagate significantly along the edge but essentially ceases to exist after about 10 inverse hoppings. 
In fact, almost the whole weight is immediately dissipated into the bulk of the system.

The data for the topologically non-trivial state in Fig.\ \ref{fig:rel} also prove that after the early dissipation of the spin-density portion carried by bulk states, the total weight of the cloud almost stays at a constant value. 
This is as expected, since after the early stage almost the whole weight of the, say, spin-up excitation is carried by the respective helical edge state only.
Furthermore, the speed of the spin-up wave packet moving to the left amounts to $v \approx 0.29$ sites per inverse hopping, as can be read off from the peak maxima.
This almost perfectly matches with the Fermi velocity $v_{\rm F} \approx 0.285$ of the edge state (see discussion of Fig.\ \ref{fig:bands}).
Interestingly, there is a low-weight peak visible for $t=20$ around site $i=43$ in the spin-up case (upper part of the figure) found to the {\em right} of the injection site $I=36$ and moving further to the right (see $t=40$).
This is due to the locally broken TRS which thus undermines spin-momentum locking to some extent.

\section{Spin injection and propagation in the presence of the read-out spin}
\label{sec:read}

\begin{figure}[b]
\includegraphics[width=0.95\columnwidth]{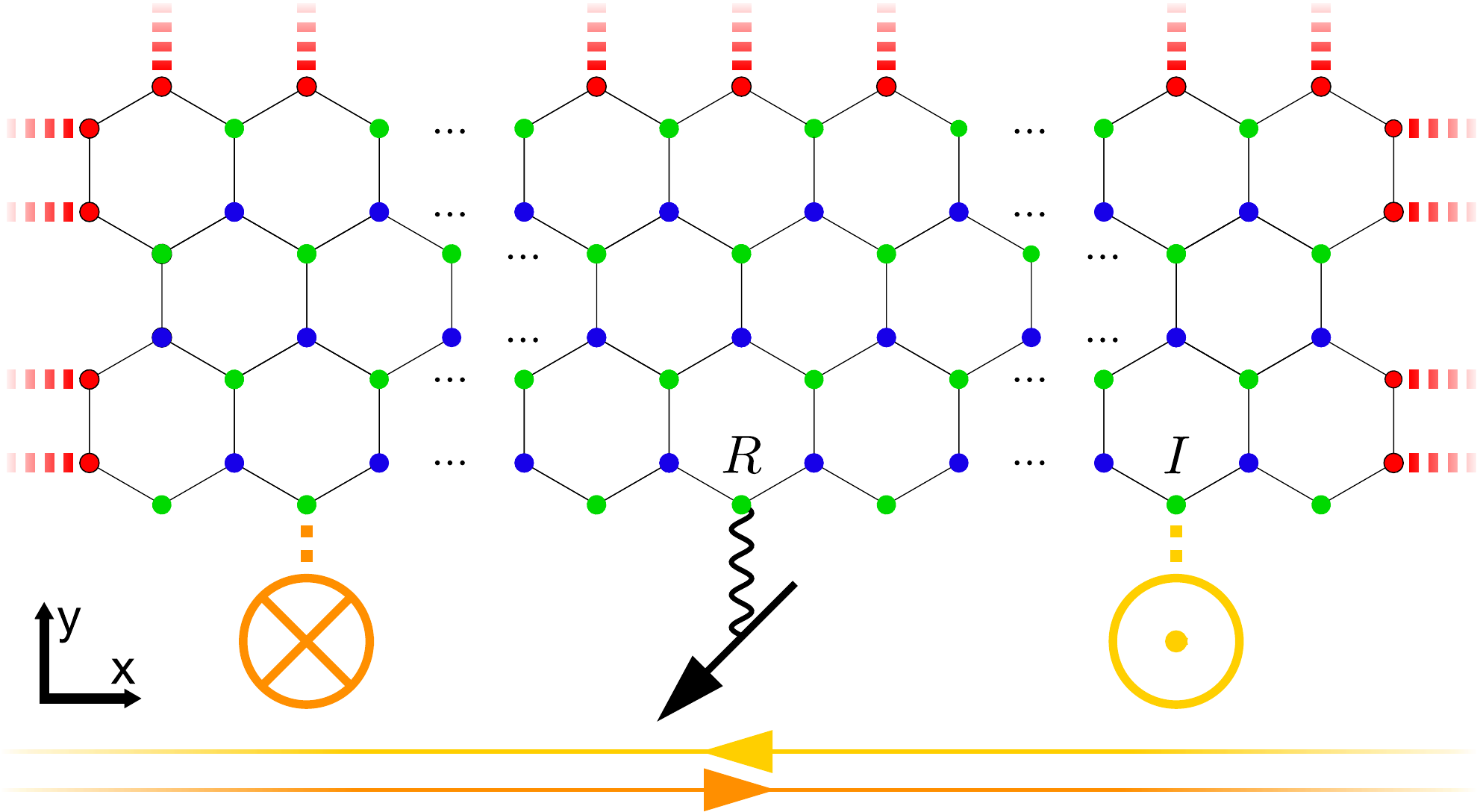}
\caption{
The same as Fig.\ \ref{fig:rib1} but with an additional classical ``read-out spin'' $\ff S_{\rm R}$ (black arrow), antiferromagnetically exchange coupled (wavy black line) to the local spin of the electron system at site $R$ of the ``physical'' zigzag edge.
$R$ is chosen to be the central site of the edge. 
The injection field $\ff B_{\rm I}$ couples to the edge site $I$ half way between $R$ and one of the corners.
In case of $\ff B_{\rm I} = + B_{\rm I} \ff e_{z}$, the induced a spin-up polarized density propagates to the left (yellow line) and scatters from $\ff S_{\rm R}$, if $I$ is located to the right of $R$.
Vice versa for a spin-down polarized density: scattering if $I$ is located to the left of $R$.
}
\label{fig:rib2}
\end{figure}

In the next step, the discussion of spin injection and propagation is extended to the full setup sketched in Fig.\ \ref{fig:rib2}. 
In addition to the local injection field, a classical read-out spin $\ff S_{\rm R}$ with $S_{\rm R} \equiv |\ff S_{\rm R}|=1/2$ is exchange coupled antiferromagnetically at the central site $R$ of the physical zigzag edge. 
We choose a generic coupling strength $J=2$ such that $J  |\ff S_{\rm R}|=1$.
The injection field is placed at a site $I$ half way between $R$ and one of the corners.

We again consider a static spin injection and fix the injection field aligned to the $z$-axis, $\ff B_{\rm I} = \pm B_{\rm I} \ff e_{z}$.
Furthermore, the initial direction of the classical spin is assumed to be in-plane, say, $\ff S_{\rm R} = S_{\rm R} \ff e_{x}$. 
The motivation of this choice is to maximize the torque exerted on the read-out spin by the propagating spin excitation that is released when switching off $\ff B_{\rm I}$.

First, we concentrate on the initial state of the electron system, which is taken to be the ground-state Fermi sea for fixed $\ff B_{\rm I}$ and $\ff S_{\rm R}$.
Such an initial state is ``stressed'', i.e., it differs from the {\em total} system ground state that also minimizes the total energy with respect to the direction of $\ff S_{\rm R}$. 
The total system ground state would be realized if both, $\ff S_{\rm R}$ and $\ff B_{\rm I}$, were lying in the $x$-$y$-plane enclosing a possibly  inter-impurity distance-dependent azimuthal angle $\Delta\phi(R)$.
Following Ref.\ \cite{GCXZ09} the angle is given by $\Delta\phi(R) = \pi -  \alpha(R)$, where $\alpha(R) = 2 R \varepsilon_{\rm F} / v_{\rm F}$ with the inter-impurity distance $R$, the Fermi energy $\varepsilon_{\rm F}$ and the Fermi velocity $v_{\rm F}$.
Note that in our case $\Delta\phi(R)$ is finite since $\varepsilon_{\rm F}=0$ should be obtained for $L\to \infty$ only.

Here, with the injection field aligned to the $z$-direction, a total-system {\em excited} state is prepared at $t=0$. 
Hence, switching off the injection field for $t>0$ to release the polarization cloud in the vicinity of site $I$ leads to a propagation of the spin density (as in Fig.\ \ref{fig:rel}), but additionally we expect an {\em immediate} dynamics of the read-out spin as well. 

\begin{figure}[t]
\includegraphics[width=0.98\columnwidth]{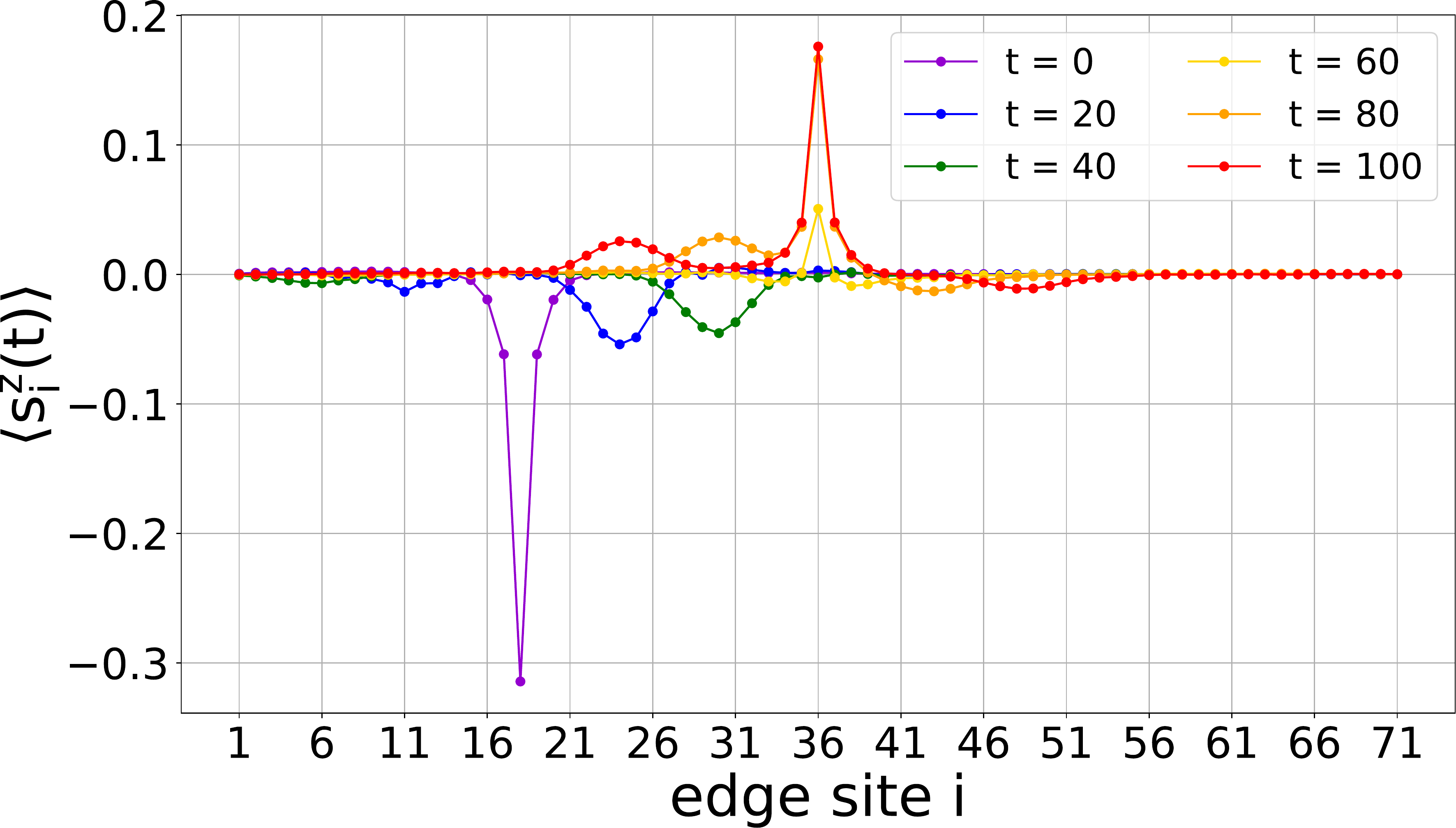}
\caption{
The same as Fig.\ \ref{fig:rel} but for a spin-down injection at site $I=18$ left to the additional classical read-out spin at site $R=36$ coupled antiferromagnetically with $J=2$. 
Initially, at time $t=0$, the electron system is in its ground state for fixed $\ff B_{\rm I} = - B_{\rm I} \ff e_{z}$ and $\ff S_{\rm R} = S_{\rm R} \ff e_{x}$.
For $t>0$, the injection field is swichted off, $\ff B_{\rm I}=0$.
Other paramters: see Fig.\ \ref{fig:rel}.
}
\label{fig:prop}
\end{figure}

Let us first concentrate on the dynamics of the injected spin-polarization cloud as shown in Fig.\ \ref{fig:prop}. 
Initially, at time $t=0$, the polarization cloud, although formed at site $I=18$ left to the read-out spin, is almost indistinguishable from the cloud initially formed at $I=36$ for the spin-down injection shown in Fig.\ \ref{fig:rel}, i.e., finite-size effects are negligible. 
For $t>0$ the spin density predominantly propagates to the right reflecting again spin-momentum locking. 
This is worth mentioning, as the presence of the read-out spin implies a broken TRS of the electronic Hamiltonian.
However, one still expects \cite{Maciejko2009} helical transport to prevail as TRS is broken only locally.
Also for $t=20$ and $t=40$ the polarization cloud does not show a sizeable perturbation caused by the read-out spin at site $R=36$. 
The snapshot for $t=60$, on the other hand, does show a strongly deformed cloud. 
Due to the locally broken TRS, the spin excitation strongly scatters from the read-out spin and is seen to back-propagate to the left for $t=80$ and $t=100$, predominantly as a {\em spin-up} excitation. 
A portion of clearly lower weight is transmitted through the classical-spin impurity and continues as a spin-down excitation to the right.

In the vicinity of the read-out spin, the polarization cloud does not change much for $t\ge 80$ (compare orange and red data points close to $R=36$).
This hints to a finite $z$-component of the read-out spin, which has trapped part of the polarization cloud close to site $R$ and whose dynamics has almost come to an end for $t=100$. 
The finite $z$-component of $\ff S_{\rm R}$ must result from the torque exerted on $\ff S_{\rm R}$ by the polarization cloud during the scattering event. 
It must further compensate the countertorque exerted by the read-out spin on the polarization cloud.

\section{Impact on the read-out spin}
\label{sec:impact}

\begin{figure}[t]
\includegraphics[width=0.7\columnwidth]{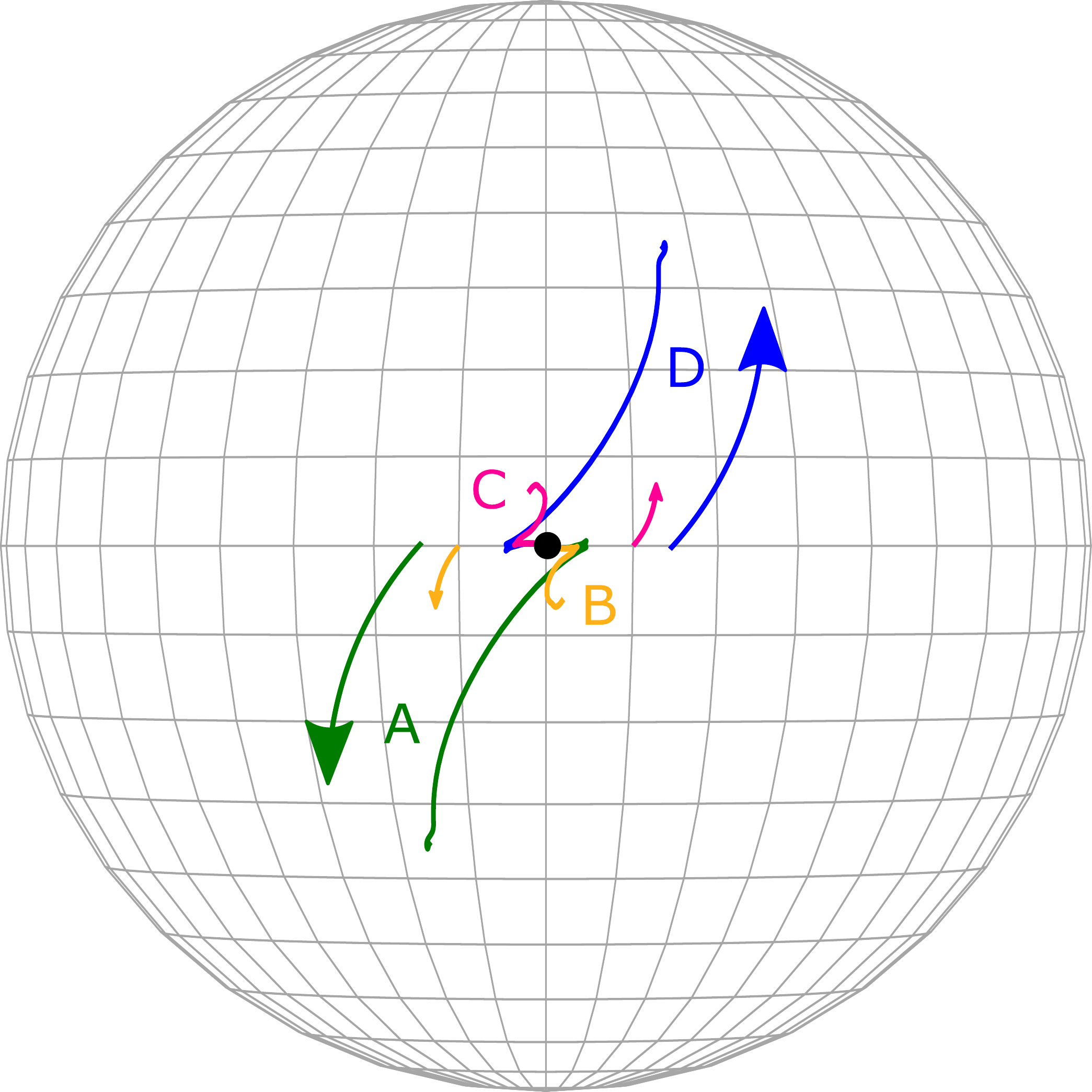}
\caption{
The path traced out by the tip of the classical read-out spin on its Bloch sphere during various processes A-D, all started with a static spin injection, switching off the injection field $\ff B_{\rm I}$ at $t=0$ and propagating the system's state up to $t=150$.
The read-out spin $\ff S_{\rm R}$ is located at the center $R=36$ of edge. 
In all cases, it points in $+x$-direction initially (black dot).
A: spin-down injection at site $I=18$ left to $R$ (``proper process'').
B: spin-down injection at site $I=54$ right to $R$ (``improper process'').
C and D: spin-up injection at $I=18$ (improper) and $I=54$ (proper), respectively. 
Parameters as in Fig.\ \ref{fig:prop}.
Colors serve to guide the eye.
The four arrows indicate the direction of the respective processes.
}
\label{fig:pi}
\end{figure}

This interpretation is in fact corroborated by looking at the dynamics of the classical spin.
Fig.\ \ref{fig:pi} displays the trajectory of the classical read-out spin on the classical ``Bloch'' sphere, starting from the point $\ff S_{\rm R}=S_{\rm R} \cdot (1,0,0)$ at $t=0$. 
The initial state of the electron system and of the injection field is prepared as above, namely a static spin injection with a finite $\ff B_{\rm I} = \pm B_{\rm I} \ff e_{z}$ and a ground-state Fermi sea.

We first discuss a spin-down injection ($\ff B_{\rm I} = - B_{\rm I} \ff e_{z}$) for two different cases: 
For the ``proper process'' A, the injection site $I$ is located to the left of the site $R$, to which the spin coupled (see Fig.\ \ref{fig:rib2}), so that the right-moving polarization cloud eventually hits the spin, while for the ``improper process'' B, $I$ is located right of $R$. 

However, even in case B, there is a non-trivial real-time dynamics after switching off $\ff B_{\rm I}$ for $t>0$. 
There are two reasons for this. 
First, the electron ground state is stressed for the chosen directions of $\ff B_{\rm I}$ and $\ff S_{\rm R}(t=0)$. 
Namely, the local spins of two neighboring edge sites enclose an angle that, depending on the distance between $I$ and $R$, weakly deviates from the angle in the electronic ground state at $\ff B_{\rm I}=0$. 
This implies a spin relaxation in the electron system, which starts immediately and leads to a finite torque on $\ff S_{\rm R}$.
However, the result is actually small. 
As can be seen in Fig.\ \ref{fig:pi} for case B, the spin moves a few degrees on the Bloch sphere and mainly within the $x$-$y$-plane only, until, at $t \approx 50$, its dynamics changes qualitatively, and a finite out-of-plane ($z$-) component builds up. 
Also in this second stage of the dynamics, the spin does not move far. 
After 150 inverse hoppings, the spin is still located close to its starting point, and there is no significant further dynamics detectable for longer propagation times. 
The cause for this second stage of the dynamics is different from the first, as this results from scattering of the low-weight portion of the polarization cloud that has started from $I$ and moved towards $R$ (while the overwhelming part of the cloud moves {\em away} from $R$, consistent with spin-momentum locking). 
This scattering event only sets in after the necessary time required to propagate from $I$ to $R$, here after $t\approx 50$ inverse hoppings. 
Its effect is weak as the weight of the excitation is small.

In case A, the proper process, on the other hand, the second stage of the dynamics must be much more effective, since now it is almost the full weight of the polarization cloud that propagates towards site $R$ and, again in a time window around $t\approx 50$, scatters from the spin. 
In fact, as can be seen in Fig.\ \ref{fig:pi}, the trajectory of the spin for case A is almost identical to that for case B at early times, where the dynamics is dominated by stress relaxation in the electron system. 
In the second stage at later times, where the spin experiences the torque of the scattering polarization cloud, is makes substantial  progress on the Bloch sphere and develops a sizable $z$-component $S_{\rm R,z} = S_{\rm R} \sin \vartheta$ with $\vartheta \approx -0.36 \cdot \pi / 2$. 

Cases C and D, obtained by starting with a spin-up injection, are perfectly symmetric to A and B, see Fig.\ \ref{fig:pi}.
Here, for $\ff B_{\rm I} = B_{\rm I} \ff e_{z}$ but for unchanged initial direction of $\ff S_{\rm R}$, the initial electron ground state is stressed with opposite helicity. 
This results in a sign change of the torque on $\ff S_{\rm R}$. 
Similarly, for the proper process D, the sign of the torque is opposite to that of case A, as now it is a spin-up cloud that scatters from the spin.

\section{Dynamic spin injection and iterating the process}
\label{sec:iter}

As the spin dynamics comes to a halt once the polarization cloud has passed by and the spin torque has diminished, the system is, at least locally at the edge, in a state close to its ground state, and thus one may start over with a second process thereafter. Some obvious questions arise in this context: 
Can we undo the rotation of the read-out spin by a subsequent second process?
Can we achieve a ``complete'' switching process $\ff S_{\rm R} \to - \ff S_{\rm R}$ by iterating the process considered so far?
Can we undo the whole iterated process?

We start the discussion with a single additional ``basic injection-pump'' (BIP) process. 
This BIP process requires (i) a spin injection, which, however, must be treated fully dynamically, and (ii) the dynamical pump of the read-out spin.

At, say, time $t=0$, the dynamic spin injection starts from an {\em arbitrary} electronic state and an {\em arbitrary} state (direction) of the read-out spin $\ff S_{\rm R}$.
One may assume, that initially the state of the electron system is the ground state (for a given $\ff S_{\rm R}$), but after the first or  after several BIP processes, the system may not have fully relaxed to its ground state before the next BIP process starts.
At $t=0$, the spin-injection field $\ff B_{\rm I} = \pm B_{\rm I} \ff e_{z}$ is switched on. 
The spin-injection process lasts up to a specified time $t=t_{\rm inj}$, when the injection field is switched off. 

\begin{figure}[b]
\includegraphics[width=0.95\columnwidth]{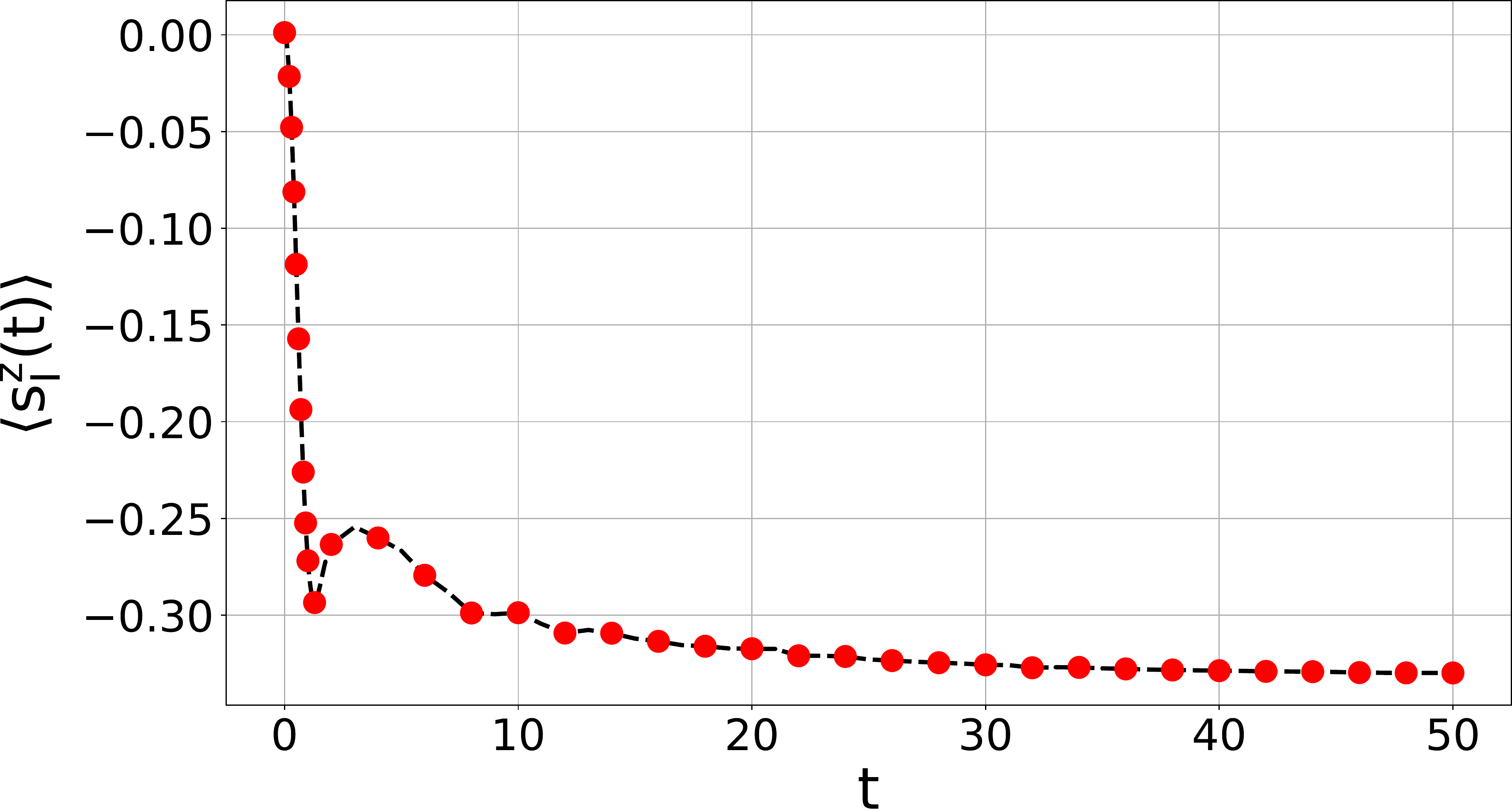}
\caption{
Real-time evolution of the $z$-component $\braket{{s}_{i}^{z}(t)}$ of the local magnetic spin moment at site $I=18$ of the physical zigzag edge during a spin-down injection process steered by an injection field $\ff B_{\rm I}$ in $-z$-direction and applied for $t_{\rm inj}=50$ inverse hoppings.
Parameters for the topologically non-trivial case as in Fig.\ \ref{fig:prop}.
}
\label{fig:inj}
\end{figure}

The magnitude of the final polarization decisively depends on the injection time $t_{\rm inj}$. 
This is demonstrated with Fig.\ \ref{fig:inj} for a spin-down injection ($\ff B_{\rm I} = - B_{\rm I} \ff e_{z}$). 
In this example the dynamical spin injection starts from the electron ground state in the presence of the classical spin pointing in $+x$-direction (see the setup shown in Fig.\ \ref{fig:rib2} with $I=18$ and $R=36$).
It shows the temporal evolution of the local magnetic spin moment $\braket{{s}_{i}^{z}(t)}$ at the injection site $i=I$. 
The results are basically independent of the choice of $I$, provided that the distance to the corners and to $R$ is large enough. 
A few sites has proven to be sufficient. 
We find that the formation of the local spin moment at $I$ (and of the entire polarization cloud) is very fast: 
After a few inverse hoppings, a considerable magnetic moment at site $I$ has formed, pointing in field direction, with a magnitude that is already close to the final saturation value.
The field strength ($B_{\rm I}=1$) is not sufficient to fully polarize the moment at $I$, i.e., $|\braket{{s}_{i}^{z}(t)}| < 0.5$. 
However, for the given strength of the field, we choose the end of the injection step such that saturation is (almost) reached. 
For $B_{\rm I}=1$ this is the case for $t=t_{\rm inj}=50$, where we get $\braket{{s}_{i}^{z}(t_{\rm pump})} \approx -0.33$.
We will stick to this injection time in the following.

The second part of the BIP process starts at $t=t_{\rm inj}$ by switching off the injection field, thus releasing the accumulated polarization cloud. 
If the location of site $I$ is properly chosen, the cloud propagates towards and scatters from the read-out spin thereby pumping the spin direction. 
The backscattered and also the transmitted part of the cloud are eventually dissipated into the Lindblad bath at the corners of the edge and partially to the bulk (and then to the Lindblad bath at other edges) simulating total-energy-conserving dissipation in a macroscopically large sample.
Finally, the pump part can be terminated at $t=t_{\rm inj}+t_{\rm pump}$, when there is no longer a significant spin dynamics, and another BIP process may follow.
A pumping time of $t_{\rm pump} = 150$ is found to be sufficient, see the discussion in Secs.\ \ref{sec:read} and \ref{sec:impact}. 
Generally, a sensible estimate for the minimal $t_{\rm pump}$ is given by the distance between sites $I$ and $R$ divided by the Fermi velocity $v_{\rm F}$ of the helical edge states.

\begin{figure}[t]
\includegraphics[width=0.7\columnwidth]{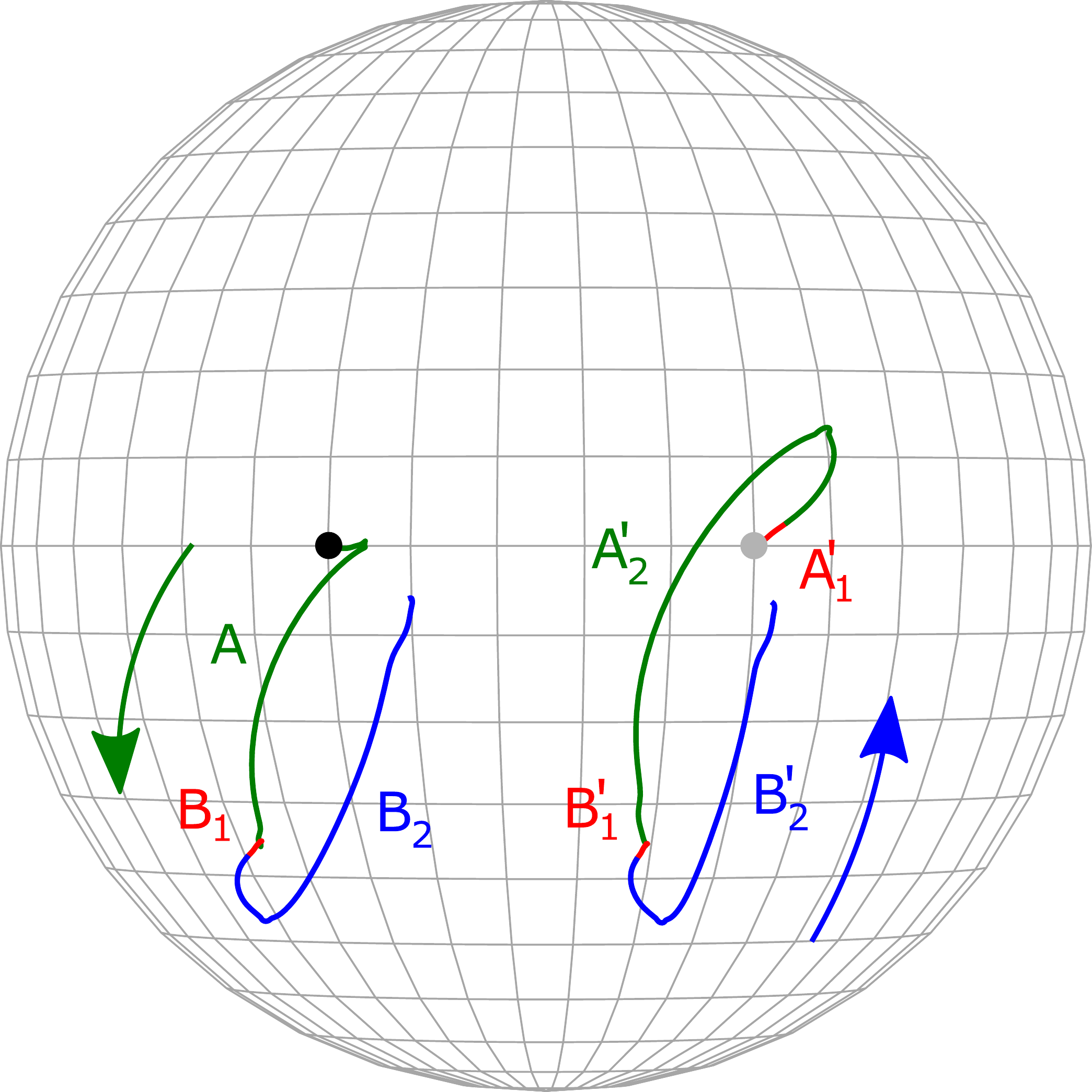}
\caption{
Process cycle on the left: 
path traced out by the tip of the classical read-out impurity spin during two concatenated processes A and B.
A (green): identical with D in Fig.\ \ref{fig:pi}, static spin-down injection, switching off the injection field at $t=0$, propagation of the system’s state up to $t_{\rm pump} = 150$, proper process with $I=54$ and $R=36$.
B: consists of B$_{1}$ and B$_{2}$, proper BIP process with $I=16$ and $R=36$.
B$_{1}$ (red): dynamic spin-up injection, $t_{\rm inj}=50$, starting from the final state of process A.
B$_{2}$ (blue): release of the injected spin-up density, propagation, scattering at $\ff S_{\rm R}$, $t_{\rm pump}=150$.
Black dot: initial read-out spin position $\ff S_{\rm R}(t=0) = S_{\rm R} \cdot (1,0,0)$.
Process cycle on the right: the same but starting, for better visibility, from 
$\ff S_{\rm R} = S_{\rm R}\cdot (1,1,0) / \sqrt{2}$ at $t=0$, see gray dot, and with A replaced by A$'$ consisting a dynamic injection A$'_{1}$ and, as before, of a pump part A$'_{2}$. 
B$'_{1}$ and B$'_{2}$: the same as B$_{1}$ and B$_{2}$ but starting from the final state of process A$'$.
}
\label{fig:2bip}
\end{figure}

The concatenated process on the left of Fig.\ \ref{fig:2bip} demonstrates that a spin-up BIP process (B), to a large extent, can in fact undo a preceding spin-down process (A).
The whole process A$+$B starts at time $t=0$ (see black dot) from a ground state in the presence of a local spin-down injection field, i.e., we consider a static spin injection. 
Switching off the field, the system evolves and the read-out spin moves towards the south pole of the Bloch sphere (green line), this is the same as process A shown in Fig.\ \ref{fig:pi}. 
Thereafter, the system is in a state, where there is essentially no more dynamics of $\ff S_{\rm R}$ and where locally, in the vicinity of sites $I$ and $R$ and the edge sites in between, the electron system has relaxed to a state that is ``close'' to the ground state, as can be monitored by tracing the temporal evolution of the local spin moments $\langle \ff s_{i} (t) \rangle$ at all sites in the nanoribbon.
After $t_{\rm pump}=150$, there are still outgoing wave packets visible that continue to be absorbed, however, by the dissipative boundary conditions. 

The spin-up BIP (process B on the left) starts from this state with a dynamical spin-up injection (first part B$_1$, red line), which lasts for 50 inverse hoppings but does not affect the orientation of the spin very much, and is completed with the second part B$_{2}$ (blue line), which lasts for another 150 inverse hopping. 
During $B_{2}$, after a short time of further downward deflection, the scattering of the polarization cloud efficiently drives the spin back towards the north pole of the Bloch sphere. 
However, it does not quite get to the original position $\ff S_{\rm R}=S_{\rm R}\cdot (1,0,0)$ at the equator but stays a little below and has drifted somewhat to the east.

There are two reasons for this ``imperfection'': 
First, processes A and B are not constructed completely inverse to each other, since in A we have employed the ``static'' injection step.
We have therefore repeated the cycle, see concatenated process on the right of Fig.\ \ref{fig:2bip}, by replacing A with a spin-down BIP process A$'$, which, as described above, consists of a {\em dynamic} injection A$'_{1}$ and a pump part A$'_{2}$ and initially starts from the ground state of the electron system for a given spin orientation of $\ff S_{\rm R}$ in the $x$-$y$-plane. 
In addition, we choose a state rotated around $\ff e_{z}$ by $\pi/4$ with $\ff S_{\rm R} = S_{\rm R} (1,1,0) / \sqrt{2}$ as a starting point, see gray dot in Fig.\ \ref{fig:2bip} on the right, just to separate in the plot the process from process A (left) on the sphere.
A$'$ is concatenated with the inverse one, i.e., with the spin-up BIP process B$'$=B$'_{1}$+B$'_{2}$.
The duration of the subprocesses of A$'$ and of B$'$ are chosen identical (50 and 150 inverse hoppings each).
As can be seen in Fig.\ \ref{fig:2bip} (right), the deviation from the starting point, which remains after completion of A$'+$B$'$, is smaller as compared to the cycle A+B on the left.

Opposed to thermodynamic processes that are steered in the state space of equilibrium or, at zero temperature, of ground states, the processes considered here involve non-equilibrium states.
However, if one would ideally extend the second (pump) part of the BIP process until the electron system is {\em fully} relaxed to its ground state, at least locally in a sufficiently extended spatial region around $I$ and $R$, the end state of process A would be {\em equivalent} to the start state of A, and one could perfectly undo A with B.
The mentioned equivalence is just a rotation.
If it was a fully relaxed, the end state of A (of $\ff S_{\rm R}$ and of the electron system as well), would merely be the start state of A, but rotated around an axis $\ff e$ and by some angle $\varphi$, where the unit vector $\ff e$ is perpendicular to the plane spanned by the center inside the Bloch sphere and the start and end state on the sphere.
Note that the electronic ground state $| \ff S_{\rm R} \rangle$ for given orientation of $\ff S_{\rm R}$ and the ground state $| \ff S'_{\rm R} \rangle$ of the model for a different orientation $\ff S'_{\rm R}$ have the same energy and that $| \ff S'_{\rm R} \rangle = U(\ff e, \varphi) | \ff S_{\rm R} \rangle$ with the according unitary rotation operator $U(\ff e, \varphi)$.
This is due to the fact that (as discussed in Sec.\ \ref{sec:rib}) the ground state of the pure Kane-Mele model (without coupling to $\ff S_{\rm R}$) is an $SU(2)$-invariant singlet, despite the SOC anisotropy.
We conclude that incomplete relaxation is exactly the second reason for the imperfect reversibility.

\begin{figure}[t]
\includegraphics[width=0.7\columnwidth]{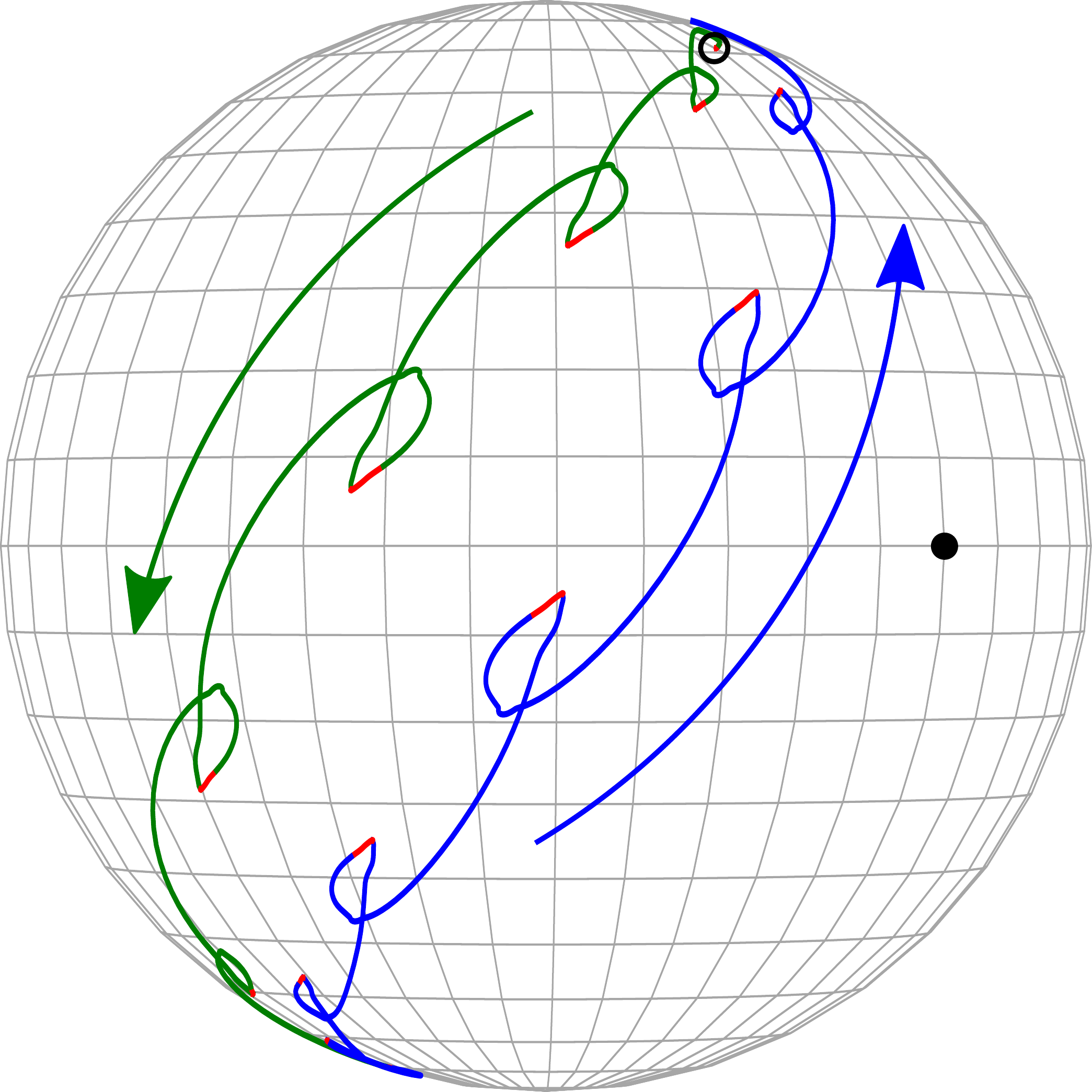}
\caption{
Trajectory of the read-out spin in a process consisting of 6 spin-down followed by 7 spin-up BIP processes starting from a state (open black dot) close to the north pole, $\eta = 0.05$, see Eq.\ (\ref{eq:eta}).
Filled black dot: position $S_{\rm R}\cdot (1,0,0)$ marking the longitude of the initial spin orientation. 
Red lines: spin injection (50 inverse hoppings).
Green lines: proper spin-down pump (150 inverse hoppings).
Blue lines: proper spin-up pump (150 inverse hoppings).
Parameters as in Figs.\ \ref{fig:rel} and \ref{fig:prop}.
}
\label{fig:nssn}
\end{figure}

\begin{figure}[t]
\includegraphics[width=0.95\columnwidth]{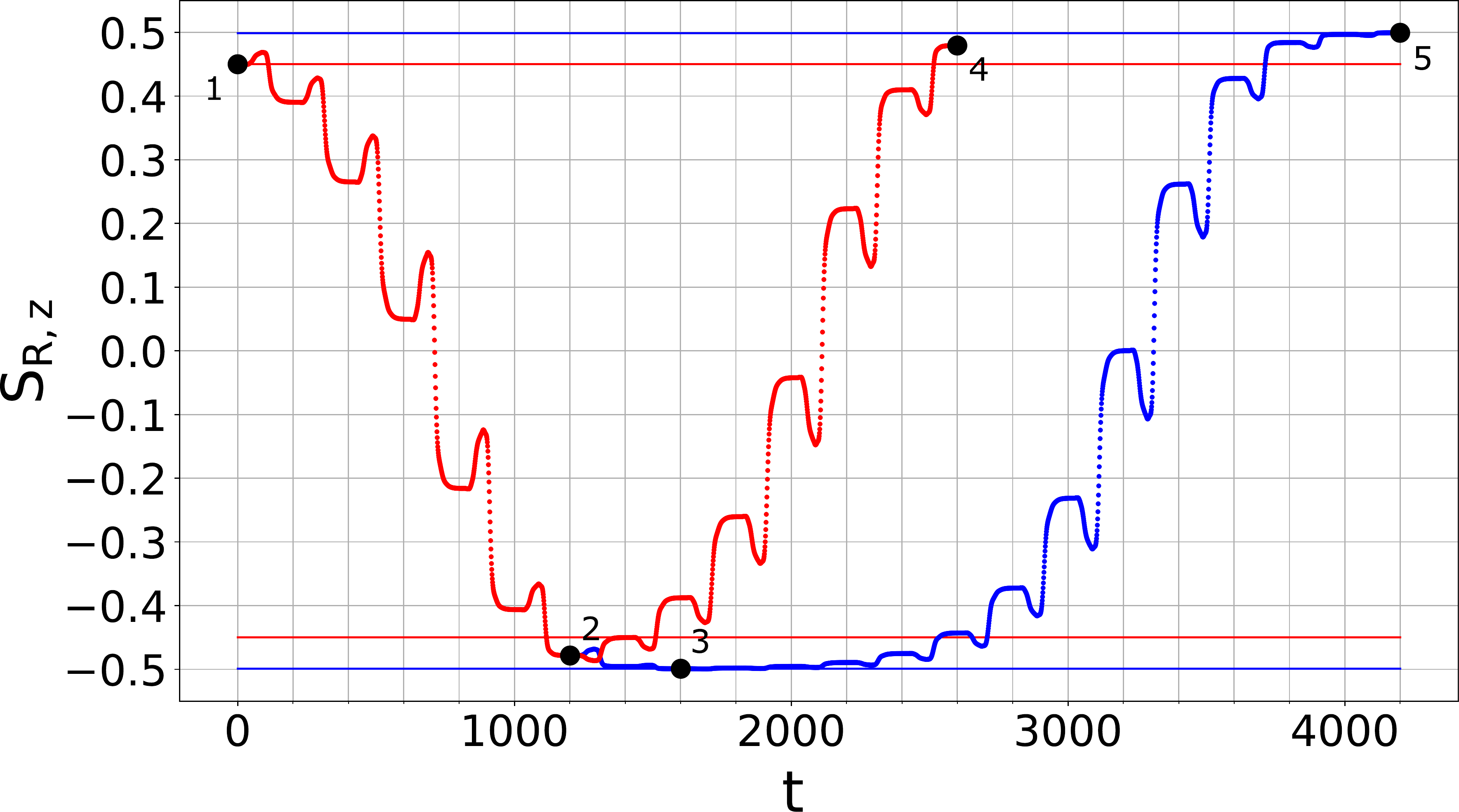}
\caption{
Temporal evolution of $S_{\rm R,z}$ corresponding to Fig.\ \ref{fig:nssn} with tolerance level $\eta=0.05$, see red data points. 
Labels 1, 2, 4 indicate start, reversal and termination of the switching cycle.
Horizontal red lines: threshold levels $\pm S_{\rm R}(1-\eta)$. 
Blue data points and lines: same starting point (1) but now $\eta=0.001$ defines reversal and termination of the process, see labels 3,5.
Vertical lines separate the different BIP processes, each lasting 200 inverse hoppings..
}
\label{fig:comp}
\end{figure}

Of course, this imperfection could be controlled by simply extending the pump part of the BIP processes. 
This, however, becomes irrelevant for the question whether a complete switching process $\ff S_{\rm R} = +S_{\rm R} \ff e_{z} \to - S_{\rm R} \ff e_{z}$ and, vice versa, a complete reversal $\ff S_{\rm R} = - S_{\rm R} \ff e_{z} \to + S_{\rm R} \ff e_{z}$ can be achieved by iterating several BIP processes.
There is one small complication to overcome.
Namely, a state $\ff S_{\rm R} = \pm S_{\rm R} \ff e_{z}$, together with the corresponding electronic ground state is non-responsive to a spin-up oder spin-down polarized wave packet passing by, since the torque on the read-out spin, $J \ff S_{\rm R} \times \langle \ff s_{\rm R} \rangle$, vanishes in this case.
We therefore start from an initial state with $\ff S_{\rm R}$ close to the north pole of the Bloch sphere but slightly tilted towards to $x$-direction:
\be
\ff S_{\rm R}(t=0) = 
S_{\rm R}
\left( 
\begin{array}{ccc}
\sqrt{1-(1-\eta)^{2}} \\ 0 \\ 1- \eta \\
\end{array}
\right) \: .
\label{eq:eta}
\ee
The parameter $\eta$ controls the deviation of the $z$-component from its maximal value in the initial state. 

Fig.\ \ref{fig:nssn} displays the path of the classical read-out spin on the Bloch sphere in an attempted full switching process. 
This process consists of 6 spin-down BIP processes, during which the $z$-component of $\ff S_{\rm R}$ is indeed found to decrease step by step in a controlled way. 
After the last spin-down BIP process, we have $S^{z}_{\rm R} < - S_{\rm R} (1-\eta )$, i.e., the spin has passed beyond the southern latitude that corresponds to the starting point on according the northern latitude, i.e., the parameter $\eta$ also serves to define a termination criterion for the switching process. 
This is necessary because adding further spin-down BIP processes would bring the spin less and less efficiently closer to the south pole.
Fig.\ \ref{fig:nssn} demonstrates that within a 5\% tolarance ($\eta=0.05$), a full switching process is in fact possible with 6 BIP processes, where each BIP process consists of 50 inverse hoppings of spin injection and 150 inverse hoppings of free time evolution, scattering and pumping the read-out spin. 

When using the same tolerance level, about the same number of BIP processes (in this case seven) are needed to subsequently reverse the switching and to bring the spin back close to the north pole of the Bloch sphere again.
It goes without saying that we need spin-up rather than spin-down BIP processes in this case. 
The near-to complete switching (green) and reversal (blue) is shown in Fig.\ \ref{fig:nssn}. 

Fig.\ \ref{fig:comp} shows the time evolution of the $z$-component of $\ff S_{\rm R}$ in detail, see red data. 
Note that each BIP process lasts for $50+150=200$ inverse hoppings as indicated by the vertical lines. 
A comparison with a process for a different, stricter tolerance level is instructive: 
With $\eta=0.001$, see blue data in the figure, but using the same starting point, the first part of the switching process is trivially the same as before, but then it takes more effort to drive the spin towards the south pole and to reach the threshold set by $\eta$. 
After reversal of the switching process, it also takes more effort, i.e., more BIP processes, to finally reach the north pole with the stricter tolerance level again.
In total 8 spin-down and 13 spin-up BIP processes are needed for a complete switching cycle. 

\section{Conclusions}
\label{sec:con}

The state of a magnetic adatom with a well-formed and stable spin moment which is exchange-coupled to the surface of a TR symmetric topological insulator can be controlled to a large extent by making use of the helical character of the topologically protected edge modes.
We have set up a simple model of a classical spin coupled to the zigzag edge of a Kane-Mele nanoribbon and studied various control protocols in detail. 
The technical key ingredient for the numerical study is the dissipative Lindblad boundary conditions imposed on all but the physically relevant edge of the ribbon. 
This has allowed us to study the coupled microscopic real-time dynamics of the spin and the electron system up to time scales of thousands of inverse hoppings without disturbing effects due to interference with propagating wave packets reflected from the boundaries of the system. 

In this study we have considered a spin-switching mechanism built from several basic injection-pump (BIP) processes, each consisting of a dynamic spin injection at a distant injection site $I$ of the edge and a subsequent pump part driving the read-out spin at site $R$.
The main idea is to exploit the topological properties of the system at all stages of the process:

(i) A local spin-up or spin-down excitation at $I$ aligned to the $z$-direction will selectively induce a spin-polarized excitation carried by the edge states. 
Their presence and their helical character is ensured by the fundamental bulk-boundary correspondence.
A local magnetic field, used here to describe the dynamical spin injection, also couples to bulk states. 
The bulk-state-supported part of the polarization cloud, however, does no longer play a role, since it is quickly dissipated into the bulk of the lattice. 
A sizable edge-state-supported spin-polarization cloud remains. 
We found that this can be built up quickly. 
After a femtosecond time scale of 50 (or more) inverse hoppings, one finds saturation of the total spin density injected locally. 
One can thus profit from a reproducible preparation step.

(ii) Switching off the coupling to the injection field releases the spin excitation, which subsequently propagates along the edge and is unidirectional thanks to the helicity of the edge states.
Its group velocity is given by the Fermi velocity of the according edge state.
During the propagation the cloud broadens but does not lose weight so that the injected spin density, apart from the dissipated bulk contribution,  entirely interacts with the read-out spin. 
As a consequence of the topological protection of the edge modes, this feature is also robust against various possible TR symmetric local perturbations that could be present in a real sample. 
Mesoscopic distances between sites $I$ and $R$ might therefore be conceivable.

(iii) The classical spin at $R$, on the other hand, does break TRS locally, and hence the polarization cloud must scatter and, again due to the topological principle of spin-momentum locking, back-propagate with essentially {\em opposite} spin polarization.
I.e., the read-out spin must exert a spin torque on the approaching polarization cloud. 
Vice versa, the spin is driven to the $\pm z$-direction due to the counter-torque of the cloud exerted on the spin.
We found that a single BIP process can change the $z$-component of the spin to some degree, depending on its initial position. 
If this pump part of the BIP process is sufficiently long, the process is revertible and one may return to the initial spin state using exactly the same spin-injection but with opposite spin direction.
Here, we found 150 inverse hoppings to be more or less sufficient to reach a fully relaxed final state that is necessary for reversibility. 
Again, we conclude that a time scale of femto- to picoseconds (or longer) is relevant for this part of the process. 

(iv) Importantly, one may concatenate various BIP processes. 
As we have demonstrated, a full switching of the spin direction between the north and the south pole can be realized within, in principle,  arbitrarily strict tolerances. 
Five to ten basic processes prove sufficient for a reasonably complete spin switching. 
Furthermore, reversing the polarization of the spin in the injection part of the BIP processes, the spin switching can be inverted as well.
Assuming a nearest-neighbor hopping scale in the range of a hundred meV, a single full switching (or back-switching) process would take place in the picosecond regime. 
However, this must be seen as a lower limit. 
Since the system approaches a fully relaxed state after each individual BIP process, the next injection step does not have to follow immediately but can be delayed.
This also means that, in principle, the intermediate relaxed states of the switching process can be controlled experimentally by techniques that do not allow for a time resolution on the pico- or sub-pico-second time scale. 

It goes without saying that the present study has a strong model character. 
When addressing real $Z_{2}$ topological materials, there are a couple of issues to be considered in addition, such as the effects of a realistic multi-band electronic structure or electron-correlation effects. 
Another highly important question is, for example, to which extent the conclusions remain valid for time-reversal-symmetric, Kondo-type magnetic impurities, which can be modeled, e.g., by quantum rather than classical spins (see, e.g., corresponding recent studies of ground-state properties \cite{AFM17,AFM19})
This question poses a formidable correlation problem to be treated in the nonequilibrium regime and in the long-time limit. 
Furthermore, it would be interesting to study the effects of imperfections regarding the structure of the material at its surface or the precise location and type of coupling of the magnetic impurities, even though the physics is expected to be largely protected by the non-trivial topology of the states involved.
For a realistic description of materials, one would also have to address the effects of magnetic anisotropies, such as the single-ion anisotropy or the Dzyaloshinskii-Moriya interaction. 
We expect that those have a considerable impact on the time scales, and further studies in this direction are well conceivable with the present computational techniques. 

\acknowledgments
This work was supported by the Deutsche Forschungsgemeinschaft (DFG) through 
the Cluster of Excellence ``Advanced Imaging of Matter'' - EXC 2056 - project ID 390715994, and through
the Sonderforschungsbereich ``Light-induced dynamics and control of correlated quantum systems'' - SFB 925 - project ID 170620586.

\appendix

\end{document}